\documentclass[aps, prd, reprint, superscriptaddress, showpacs, nofootinbib, nobibnotes, longbibliography, amsmath, amssymb]{revtex4-1}

\usepackage[utf8]{inputenc}
\usepackage[english]{babel}
\usepackage{graphicx}
\usepackage[range-phrase = -]{siunitx}
\usepackage[caption=false]{subfig}
\usepackage{xpatch}
\usepackage{hyperref}

\usepackage{color}
\definecolor{darkblue}{cmyk}{1.0,1.0,0,0}
\hypersetup{colorlinks = true, allcolors = darkblue}

\DeclareDocumentCommand\d{}{\operatorname{d}\!}
\newcommand{\xvec}{\mathbf{x}}
\newcommand{\yvec}{\mathbf{y}}
\newcommand{\pvec}{\mathbf{p}}
\newcommand{\pabs}{\lvert\pvec\rvert}
\newcommand{\Ftrans}{F_\perp}
\newcommand{\Flong}{F_\parallel}
\newcommand{\rhotrans}{\rho_\perp}
\newcommand{\rholong}{\rho_\parallel}

\newcommand{\coloreqref}[1]{\textcolor{darkblue}{\eqref{#1}}}
\newcommand{\colorcite}[1]{\textcolor{darkblue}{\cite{#1}}}

\begin{document}

\title{\texorpdfstring{Nonthermal Fixed Points in Quantum Field Theory\\Beyond the Weak-Coupling Limit}{Nonthermal Fixed Points in Quantum Field Theory Beyond the Weak-Coupling Limit}}

\author{J\"urgen Berges}
\affiliation{ITP, Universit\"at Heidelberg, Philosophenweg 16, 69120 Heidelberg, Germany}
\author{Benjamin Wallisch}
\email{b.wallisch@damtp.cam.ac.uk}
\affiliation{ITP, Universit\"at Heidelberg, Philosophenweg 16, 69120 Heidelberg, Germany}
\affiliation{DAMTP, University of Cambridge, Wilberforce Road, Cambridge CB3 0WA, United Kingdom}

\begin{abstract}
Quantum systems in extreme conditions can exhibit universal behavior far from equilibrium associated to nonthermal fixed points with a wide range of topical applications from early-universe inflaton dynamics and heavy-ion collisions to strong quenches in ultracold quantum gases. So far, most studies have relied on a mapping of the quantum dynamics onto a classical-statistical theory that can be simulated on a computer. However, the mapping is based on a weak-coupling limit, while phenomenological applications often require moderate interaction strengths. We report on the observation of nonthermal fixed points directly in quantum field theory beyond the weak-coupling limit. For the example of a relativistic scalar $\mathrm{O}(N)$-symmetric quantum field theory, we numerically solve the nonequilibrium dynamics employing a $1/N$ expansion to next-to-leading order, which does not rely on a small coupling parameter. Starting from two different sets of overoccupied and of strong-field initial conditions, we find that nonthermal fixed points are not restricted to parameter ranges suitable for classical-statistical simulations, but extend also to couplings of order one. While the infrared behavior is found to be insensitive to the differences in the initial conditions, we demonstrate that transport phenomena to higher momenta depend on the presence or absence of a symmetry-breaking field expectation value.   
\end{abstract}

%\pacs{11.10.Wx, 67.85.De, 25.75.-q, 98.80.Cq}

\maketitle

\section{\label{sec:Introduction}Introduction and Overview}

The existence of transient universal regimes, where even quantitative agreements between seemingly disparate physical systems can be observed, drives a remarkable convergence of research activities across traditional lines of specialization. An important class of universal scaling phenomena occurs in extreme conditions far from equilibrium, characterized by unusually strong fields or large occupancies of characteristic modes when compared to thermal equilibrium with the same energy density. Such a transient overoccupation may be found in a variety of physical applications on vastly different energy scales, ranging from (p)reheating dynamics in the early universe~\colorcite{Khlebnikov:1996mc, Micha:2002ey, Berges:2008wm} to the initial stages of relativistic heavy-ion collisions~\colorcite{Lappi:2006fp, Gelis:2010nm, Berges:2013eia, Berges:2013fga}, and strong quenches in ultracold quantum gas experiments~\colorcite{Scheppach:2009wu, Nowak:2011sk, Orioli:2015dxa}. Important universality classes out of equilibrium have been recently discovered in these cases, providing exciting new links between the dynamics of cold gases and hot plasmas~\colorcite{Berges:2014bba}.

So far, most quantitative results rely on the presence of a sufficiently weak coupling parameter. In this case, essential aspects of the many-body quantum dynamics of strong fields or highly occupied modes can be mapped onto a classical-statistical field theory problem~\colorcite{Son:1996uv, Khlebnikov:1996mc, Aarts:2001yn}, which can be solved on a computer. Since universal phenomena are insensitive to details of the underlying system parameters, such as values of couplings or precise initial conditions for the subsequent evolution, one may hope that similar results persist beyond the weak-coupling limit. A corresponding fact is well known for universal scaling properties in thermal equilibrium near phase transitions associated to renormalization group fixed points, where systems with very different microscopic interaction strengths exhibit common macroscopic properties. Whether such a statement can also be made for universal scaling phenomena far from equilibrium near nonthermal fixed points, associated to turbulent transport phenomena, is a question of utmost phenomenological relevance. First indications come from holographic strong-coupling methods applied to superfluids~\colorcite{Adams:2012pj, Ewerz:2014tua}. It is the main objective of this work to provide an explicit example, where this question can be answered directly in quantum field theory beyond the weak-coupling limit.  

\vskip4pt
Motivated by the phenomenology of heavy-ion collision experiments, where the relevant gauge coupling at early times is expected to be neither particularly small nor large, there is a series of studies trying to understand extreme conditions far from equilibrium beyond the limit of weak interactions. Classical field simulations have been employed in~\colorcite{Epelbaum:2011pc, Dusling:2012ig, Gelis:2013rba} for gauge field and scalar field theories at larger couplings. They seem to demonstrate the absence of any transient dynamical scaling regime by showing a rapid approach to a thermal-like Rayleigh-Jeans distribution~\colorcite{Epelbaum:2011pc} or fast isotropization for expanding systems~\colorcite{Dusling:2012ig, Gelis:2013rba}. However, it has been shown in~\colorcite{Berges:2013lsa, Berges:2014yta} that one exceeds the range of validity of the classical descriptions for the larger couplings employed: a spurious decay of the quantum vacuum into propagating particle modes is observed, which then dominate the thermalization or isotropization dynamics.

Another approach employs effective kinetic descriptions. Kinetic theory can describe the long-time behavior of quantum systems if the typical occupancies are not too large and if their typical momenta are much larger than the in-medium screening scale. While the applicability of the perturbative kinetic theory requires a sufficiently weak coupling parameter, recent extrapolations beyond the weak-coupling limit in the context of heavy-ion collisions achieve a remarkable consistency with phenomenological expectations~\colorcite{Kurkela:2015qoa}. If applied to stronger interaction strengths, the kinetic descriptions show significant deviations from the known classical-statistical scaling behavior in the weak-coupling limit for gauge fields~\colorcite{Kurkela:2015qoa} as well as scalar field theory~\colorcite{Epelbaum:2015vxa} because of the additional quantum corrections that now become of the same order as the classical contributions in the perturbative description. It remains an important open question whether these corrections are reliably estimated if applied at stronger couplings.\footnote{For the extrapolation to stronger couplings, it would be important to resolve also some puzzling questions that remain even in the weak-coupling limit since kinetic descriptions for longitudinally expanding systems apparently fail to reproduce some relevant results of classical-statistical simulations, such as the ratio of longitudinal to transverse pressure characterizing isotropization for longitudinally expanding systems in the regime where both are expected to have an overlapping range of validity~\colorcite{Berges:2015ixa}.} 

\vskip4pt
In order to be able to judge the validity of extrapolations beyond the weak-coupling limit, a description of the dynamics based on a nonperturbative expansion parameter would be extremely valuable. For $N$-component scalar quantum field theories, this is indeed possible based on a systematic large-$N$ expansion, which has a long history for the description of scaling phenomena. At next-to-leading order (NLO) in a resummed large-$N$ expansion based on the two-particle irreducible (2PI) effective action~\colorcite{Berges:2001fi, Aarts:2002dj}, the approach is already known to analytically describe well the self-similar scaling solutions of nonthermal fixed points in the weak-coupling limit~\colorcite{Orioli:2015dxa, Berges:2015kfa}. However, to extend to stronger interactions a numerical evaluation seems mandatory. In particular, we are not only interested in the scaling solution itself, but also want to answer the question for what range of couplings a nonthermal fixed point is approached from generic strong-field or overoccupied initial conditions. While the nonthermal fixed point is known to have attractor properties in the limit of weak couplings, such that no fine-tuning of initial conditions or parameters is required to approach it, this is not guaranteed at stronger couplings even if the fixed point solution itself would still exist.    

In this work, we present a numerical solution of the time evolution equations at NLO in the $1/N$ expansion of a self-interacting $\mathrm{O}(N)$-symmetric quantum field theory in $3+1$ space-time dimensions without expansion. Because of the non-Markovian nature of the evolution equations, it is very demanding to go beyond the earlier times considered in previous studies~\colorcite{Berges:2008wm}. This is achieved in this work through algorithmic and computational advances, which allow us for the first time to study the approach to nonthermal fixed points and the subsequent deviation evolution towards thermal equilibrium. The latter is not accessible in classical-statistical descriptions even at weak couplings because of the Rayleigh-Jeans divergence of classical field theory. 

One of the main results of this work is that even for couplings of order one we find the transient universal scaling results as observed before from classical-statistical simulations in the weak-coupling limit. Remarkably, the scaling behavior for infrared momentum modes is found to persist even for much stronger interaction strengths, while the lower-occupied higher momentum modes deviate from scaling in this case. In general, scaling is seen to hold only in momentum regimes with modes having occupancies larger than one. Once the characteristic occupancies drop below one at later times, the approach to a Bose-Einstein distribution is observed as expected. Although our results on the presence and relevance of nonthermal fixed points beyond the weak-coupling limit seem to be in conflict with extrapolated kinetic theory expectations, our findings are in line with corresponding facts known from universal scaling phenomena in thermal equilibrium. 

\vskip4pt
This paper is organized as follows. In Sec.~\ref{sec:Theory}, we discuss the evolution equations beyond the weak-coupling limit and the employed initial conditions. Sections~\ref{sec:FluctuationIC} and~\ref{sec:FieldIC} cover the results for (a)~overoccupied and (b)~strong-field initial conditions, respectively. It is shown that the particle cascade towards the infrared is insensitive to the different initial conditions employed, while the energy cascade towards higher momenta depends on the presence or absence of a symmetry-breaking field expectation value. The energy cascade for the symmetric case with a vanishing field expectation value is observed here to emerge from the nonequilibrium evolution for the first time. We also compare to the classical-statistical field theory limit and demonstrate the decay of the quantum-half for strong enough couplings as an artifact of classical approximations applied beyond their range of validity. Section~\ref{sec:Summary} concludes with a summary. In a set of appendices, we review the equations of motion from the $1/N$ expansion of the 2PI effective action (App.~\ref{app:2piEoM}), describe the performed mass renormalization (App.~\ref{app:renormalization}) and provide the energy-momentum tensor (App.~\ref{app:energyMomentumTensor}) for completeness.

\section{\label{sec:Theory}Evolution Equations Beyond the Weak-Coupling Limit}

\subsection[Large-\texorpdfstring{$N$}{N} Expansion to NLO]{\label{sec:largeN_NLO}Large-N Expansion to NLO}

We consider a relativistic $\mathrm{O}(N)$-symmetric real $N$-component scalar field theory with quartic self-interaction in $3+1$-dimensional Minkowski space-time. The $N$-component field theory often serves in paradigmatic models to describe early-universe inflaton dynamics. For $N=4$, the symmetry group employed is locally isomorphic to a chiral $\mathrm{SU}_{\hskip-1pt L}(2) \times \mathrm{SU}_{\hskip-1pt R}(2)$ symmetry relevant in the context of low-energy effective descriptions of quantum chromodynamics. In addition, the $\mathrm{O}(4)$ model also reflects the field content of the Higgs sector of the Standard Model of particle physics. Since we anticipate nonrelativistic dynamics at sufficiently low momenta~\colorcite{Orioli:2015dxa}, there is also a direct correspondence to quantum Gross-Pitaevskii models for a single charged Bose field for $N=2$, for example.

\vskip4pt
As we aim to go beyond the weak-coupling limit, we take into account quantum corrections according to their scaling with the number of field components $N$. Here, $1/N$ is the (small) expansion parameter, which is not an expansion in powers of the coupling parameter. As a consequence, the $1/N$ expansion is not restricted to weak interactions in general and has been successfully employed also in nonperturbative contexts, such as the description of critical phenomena~\colorcite{ZinnJustin:2002ru}. Moreover, the resummed $1/N$ expansion based on the 2PI effective action employed is known to exhibit improved convergence properties for the computation of thermal scaling phenomena~\colorcite{Alford:2004jj} and the resummation of secular terms required to describe the long-time behavior of quantum fields out of equilibrium~\colorcite{Berges:2001fi}.   

The classical Lagrangian density for the massless field theory reads\hskip1pt\footnote{We use the Minkowski metric with signature $(1,-1,-1,-1)$, the notation $x=(x^0,\xvec)$ for time $x^0=t$ and spatial coordinates $\xvec$, and units where the speed of light, the reduced Planck constant and Boltzmann's constant equal unity, $c=\hbar=k_\mathrm{B}=1$.}
\begin{equation}
\mathcal{L}[\varphi]= \frac{1}{2}\partial_\mu \varphi_a \partial^\mu \varphi_a - \frac{\lambda}{4! N}\big(\varphi_a \varphi_a\big)^{\!2}
\label{eq:Lagrangian}
\end{equation}
with the field components $\varphi_a(x)$ and coupling parameter $\lambda$. Summation over repeated field space indices $a=1,\ldots,N$ as well as Lorentz indices $\mu=0,1,2,3$ is implied. For the corresponding classical equations of motion,	
\begin{equation}
\left(\partial_\mu \partial^\mu + \frac{\lambda}{6N} \varphi^2(x)\right)\varphi_a(x) = 0 \, 
\end{equation}	
with $\varphi^2 \equiv \varphi_a \varphi_a$, one can always scale out the dependence on the coupling by introducing a rescaled field,\footnote{The same can be done also for a massive classical scalar field.} 
\begin{equation}
\varphi_a(x) \, \rightarrow \, \frac{1}{\sqrt{\lambda}}\,  \varphi_a(x)  \, .
\label{eq:fieldRescaling}
\end{equation}
The rescaled classical field obeys the evolution equation
\begin{equation}
\left(\partial_\mu \partial^\mu + \frac{1}{6N}\varphi^2(x)\right)\varphi_a(x) = 0 \, 
\label{eq:rescaledEoM}
\end{equation}			
and, as a consequence, in classical(-statistical) dynamics any dependence on the coupling only enters via the initial conditions for the solution of the differential equation~\coloreqref{eq:rescaledEoM}. 

In contrast, the coupling cannot be scaled out from the corresponding quantum dynamics, where the classical field $\varphi_a(x)$ is replaced by a Heisenberg field operator $\hat{\varphi}_a(x)$. The quantum description is crucial, for example, to overcome the Rayleigh-Jeans divergence of classical-statistical field theory~\colorcite{Berges:2001fi}.

However, simple approximations may not reflect this distinctive property of the quantum theory. For instance, at leading order (LO) in the expansion in powers of $1/N$, the quantum expectation value of the field operator,
\begin{equation}
\phi_a(x) \equiv \langle \hat{\varphi}_a(x) \rangle \, ,
\label{eq:macroscopicField}
\end{equation}
obeys---after a rescaling corresponding to~\coloreqref{eq:fieldRescaling}---the evolution equation~\colorcite{Aarts:2002dj}
\begin{equation}
\left(\partial_\mu \partial^\mu + \frac{1}{6N}\left[\phi^2(x) + F(x,x)\right]\right)\phi_a(x) = 0
\label{eq:EoMlargeN}
\end{equation}
with $\phi^2 \equiv \phi_a \phi_a$. In comparison to the classical evolution equation~\coloreqref{eq:rescaledEoM}, the extra term $F(x,x)\equiv F_{aa}(x,x)$ in~\coloreqref{eq:EoMlargeN} stems from the connected part of the \textit{anti-commutator} expectation value of two Heisenberg field operators,
\begin{equation}
F_{ab}(x,y) \equiv \frac{1}{2} \left\langle \left\{\hat{\varphi}_a(x), \hat{\varphi}_b(y) \right\}\right\rangle - \phi_a(x) \phi_b(y) \, .
\label{eq:defF}
\end{equation}  
This anti-commutator or so-called statistical two-point function obeys at LO a similar equation as the field expectation value~\colorcite{Aarts:2002dj}:
\begin{equation}
\left(\partial_\mu \partial^\mu + \frac{1}{6N}\left[\phi^2(x) + F(x,x)\right]\right) F(x,y) = 0 \, .
\end{equation}
Accordingly, no dependence on the coupling parameter $\lambda$ appears in the evolution equations for the rescaled quantum field at LO in a large-$N$ expansion.

This changes at NLO in the $1/N$ expansion, where the evolution equations start to depend explicitly on the \textit{commutator} expectation value~\colorcite{Berges:2001fi, Aarts:2002dj}
\begin{equation}
\rho_{ab}(x,y) = \mathrm{i} \left\langle\left[ \hat{\varphi}_a(x),\hat{\varphi}_b(y)\right] \right\rangle \, .
\label{eq:defRho}
\end{equation}
The spectral function~\coloreqref{eq:defRho} encodes the equal-time commutation relation of the quantum theory,
\begin{equation}
\partial_{x^0} \rho_{ab}(x,y)|_{x^0=y^0} = \delta_{ab}\, \delta(\xvec - \yvec) \, ,
\end{equation}
and the somewhat lengthy evolution equations at NLO are given in App.~\ref{app:2piEoM} in detail. 

\begin{figure}
\vspace{-1mm}
\includegraphics[width=\columnwidth,trim={3mm 0 2mm 3mm},clip]{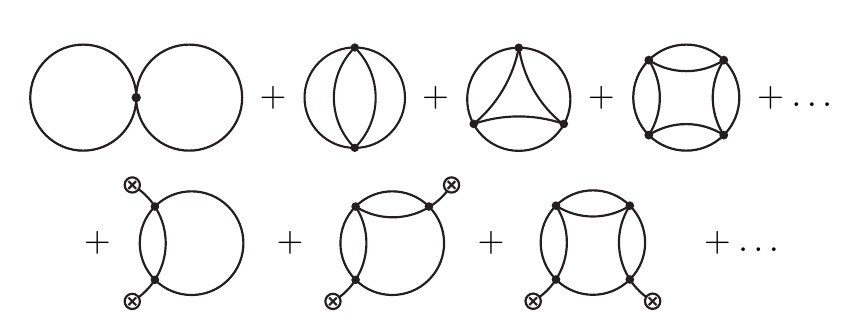}
\caption{\label{fig:diagrams}Diagrammatic representation of the infinite series of contributions to the free energy density at NLO in the \mbox{$1/N$ expansion} of the 2PI effective action. The lines denote self-consistent propagators, solid circles denote vertices and crosses denote field expectation values.}
\end{figure}

For the purpose of the present discussion, the characteristic dependence on the coupling parameter entering at NLO can be illustrated by the ``one-loop'' self-energy term
\begin{equation}
\Pi_F(x,y) = \frac{1}{6N}\left[F^2(x,y) - \left(\frac{\lambda}{2}\right)^{\!2} \rho^2(x,y)\right] \, 
\label{eq:oneLoopSelfEnergy}
\end{equation}
with $F^2 \equiv F_{ab}F_{ab}$ and $\rho^2 \equiv \rho_{ab}\rho_{ab}$. This self-energy term represents an essential building block of the infinite number of one-loop-type ``ring diagrams'' appearing at NLO in the large-$N$ expansion, as indicated for the example of the free energy density or 2PI effective action in Fig.~\ref{fig:diagrams}.

The commutator term $(\lambda/2)^2 \rho^2$ in~\coloreqref{eq:oneLoopSelfEnergy} encodes the genuine coupling dependence entering the NLO evolution equations. In fact, implementing the $1/N$ expansion to NLO in the corresponding classical-statistical theory, one can check that the commutator term is absent classically, while the $F^2$-term in~\coloreqref{eq:oneLoopSelfEnergy} has a classical analogue~\colorcite{Aarts:2001yn}. This reflects the fact that classical-statistical approximations for the description of quantum dynamics require sufficiently weak couplings, such that contributions from the $(\lambda/2)^2 \rho^2$-term are sufficiently small compared to those from the $F^2$-term. Accordingly, by solving the quantum evolution equations, one expects to confirm characteristic results known from classical-statistical simulations in the weak-coupling limit for the same initial conditions, while this may not be expected at sufficiently large values of $\lambda$.

\subsection{\label{sec:initialConditions}Initial Conditions}

We consider spatially homogeneous and isotropic initial conditions. As a consequence, the macroscopic field~\coloreqref{eq:macroscopicField} only depends on time $t$, $\phi_a(t)$, and the correlation functions depend on times and relative spatial coordinates, $F_{ab}(t,t^\prime,\lvert\xvec-\yvec\rvert)$ and $\rho_{ab}(t,t^\prime,\lvert\xvec-\yvec\rvert)$. The initial macroscopic field is chosen to point in a given direction in field index space as $\phi_a=\phi\,\delta_{a1}$, which is preserved by the nonequilibrium dynamics of the $\mathrm{O}(N)$-symmetric quantum theory. Accordingly, we can take the two-point functions to be diagonal with one longitudinal and $N-1$ transverse parts, i.e.\ $(F_{ab}) = \mathrm{diag}\!\left\{\Flong,\Ftrans,\ldots,\Ftrans\right\}$ and $(\rho_{ab}) = \mathrm{diag}\!\left\{\rholong,\rhotrans,\ldots,\rhotrans\right\}$.

The initial size of the rescaled macroscopic field, $\phi_0\equiv\phi(t=0)$, will be characterized by $\sigma_0$ defined as
\begin{equation}
\sigma_0 = \frac{\phi_0}{\sqrt{6 N}} \, ,
\end{equation}
while the time derivative of the field is taken to vanish at initial time $t=0$. Apart from initial conditions with a large macroscopic field, we consider the case of an initially highly occupied gas of free particles with characteristic (highest) momentum~$Q$. More precisely, a class of initial conditions for the anti-commutator two-point function will be described in spatial Fourier space with three-momentum $\pvec$ as
\begin{equation}
F_{\parallel,\perp}(t=t^\prime = 0,\pabs) = \frac{n_0 \Theta(Q-\pabs) + \lambda/2}{\omega_{\parallel,\perp}(0,\pabs)}
\label{eq:initialConditionsF}
\end{equation}
with occupancy parameter~$n_0$, Heaviside step  function~$\Theta$ and initial dispersion
\begin{equation}
\omega_{\parallel,\perp}(0,\pabs) = \sqrt{\pvec^2 + M^2_{\parallel,\perp}\!(0)} \, .
\label{eq:initialDispersion}
\end{equation}
Here, $M_{\parallel,\perp}\!(0)$ denotes the initial renormalized in-medium mass of longitudinal and transverse modes, respectively (cf.~App.~\ref{app:renormalization}). 

Taking into account that we have performed a field rescaling according to~\coloreqref{eq:fieldRescaling}, the actual initial occupancy per mode, $f(t=0, \pabs)$, is given by
\begin{equation}
f(t=0, \pabs) = \frac{n_0}{\lambda}\, , \quad \pabs \le Q
\label{eq:initialf}
\end{equation} 
according to~\coloreqref{eq:initialConditionsF}. This initial occupancy is supplemented by the quantum-half which appears as $\lambda/2$ in the initial correlator~\coloreqref{eq:initialConditionsF} for the rescaled field. For large typical occupancy $n_0/\lambda$, the function~\coloreqref{eq:initialf} clearly represents a far-from-equilibrium distribution, in sharp contrast to a thermal (Bose-Einstein) distribution where the characteristic occupancy is $\mathcal{O}(1)$ for momenta of order the temperature. 

Neglecting $M_{\parallel,\perp}\!(0)$ for a moment, the total initial energy density $\epsilon_0$ is approximately given by
\begin{equation}
\epsilon_0 \simeq \frac{3}{2} \frac{N}{\lambda} \sigma_0^4 + \frac{n_0}{8 \pi^2} \frac{N}{\lambda} Q^4 \, .
\end{equation}
In contrast to the contribution \mbox{$\sim N \sigma_0^4/\lambda$} from the macroscopic field, the relativistic gas contribution \mbox{$\sim N Q^4/\lambda$} comes with a significant suppression factor $(8 \pi^2)^{-1}$ unless we consider initial occupancy parameter values of order $n_0=100$. In the following, the latter value will be employed for $n_0$, which is also in line with similar choices in studies employing classical-statistical lattice simulations (see e.g.~\colorcite{Orioli:2015dxa}). 

\begin{figure}
\vspace{-0.3cm}
\subfloat[\label{fig:fluctuationInitialConditions}Fluctuation initial condition.]{%
\includegraphics*[width=0.49\columnwidth]{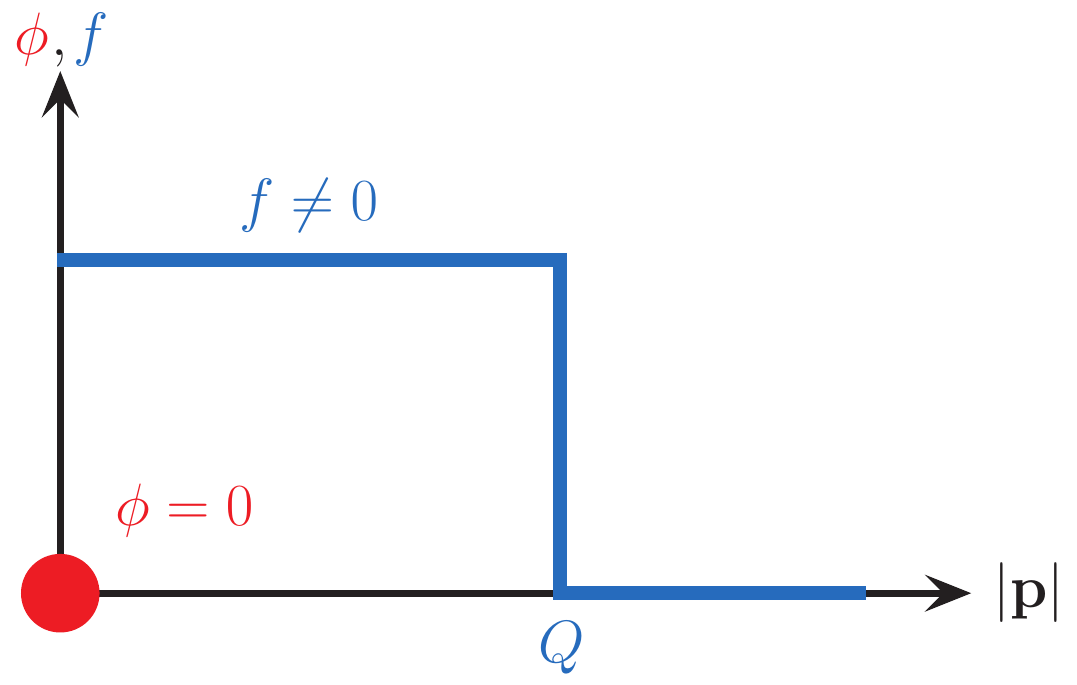}%
}
\subfloat[\label{fig:macroscopicFieldInitialConditions}Macroscopic field initial condition.]{%
\includegraphics*[width=0.49\columnwidth]{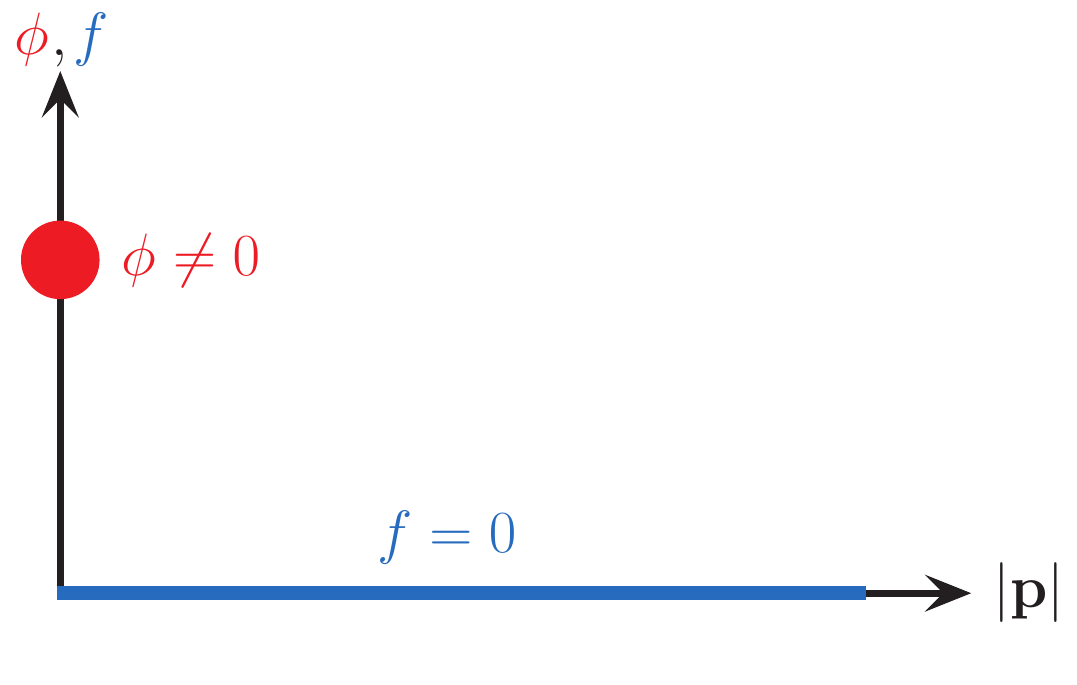}%
}\hfill
\caption{\label{fig:initialConditions}Illustration of the employed initial condition scenarios.\vspace{-2pt}}
\end{figure}

\vskip4pt
We will consider two generic types of initial condition scenarios here, which are frequently considered in the literature: (a)~fluctuation initial conditions with~$n_0 \neq 0$ and~$\sigma_0 = 0$, and (b)~macroscopic field initial conditions with~$\sigma_0 \neq 0$ in the presence of vacuum fluctuations, i.e.\ with~$n_0 = 0$. This is schematically depicted in Fig.~\ref{fig:initialConditions}.%

The macroscopic field initial condition~(b), which is also referred to as the strong-field initial condition, is known to lead to the phenomenon of parametric resonance at early times, which has been studied by the techniques employed here in~\colorcite{Berges:2002cz}. The parametric resonance instability quickly leads to large occupancies in a characteristic momentum range, similar to what is considered with the fluctuation initial condition~(a), which is also known as the overoccupied initial condition. However, there is an important difference: While in scenario (b) the presence of a nonzero field expectation value singles out an initial direction in field index space such that the $\mathrm{O}(N)$ symmetry is not manifest, scenario (a) proceeds entirely in the symmetric regime because the initially vanishing field expectation value and time derivative remain identically zero at all times by virtue of the $\mathrm{O}(N)$ symmetry. In particular, a nonzero macroscopic field is known to induce effective cubic self-interactions even though the Lagrangian~\coloreqref{eq:Lagrangian} only exhibits quartic interactions. We will analyze the impact of the underlying different scattering processes for scenarios~(a) and~(b) in the subsequent Secs.~\ref{sec:FluctuationIC} and~\ref{sec:FieldIC}, respectively.  

\vskip4pt
We consider the time evolution of systems that start far away from thermal equilibrium. Stated differently, initially---or after a characteristic short time starting from macroscopic field initial conditions~\colorcite{Berges:2002cz}---the typical occupancy is much larger than in thermal equilibrium, i.e.~$f \sim n_0/\lambda \gg 1$. For $n_0 = 100$, this already limits the maximum value of the coupling parameter to be $\lambda\lesssim\mathcal{O}(10)$, which we will not exceed in our study. In turn, starting far from equilibrium with $n_0/\lambda \gg 1/2$, the quantum corrections are expected to play no significant role at initial times. In contrast, quantum corrections become relevant at later times when the typical occupancy decreases as the system evolves on its way towards thermal equilibrium. The stronger the coupling, the earlier these quantum corrections are expected to set in. Here, an important question is to find out up to what coupling strength the system still approaches a transient nonthermal fixed point, as is known to occur in the weak-coupling limit, before the quantum corrections lead to thermal equilibrium asymptotically. 

Numerical results will be presented for $N=4$, and we will cover a coupling range of $10^{-2} \leq \lambda \leq 10$ for a cutoff-regularized theory. More precisely, we perform a self-consistent mass renormalization at initial time as detailed in App.~\ref{app:renormalization} and verify that the results are insensitive to variations of the employed ultraviolet momentum cutoff. For the field initial condition, we take $M^2_{\parallel}(0)/\sigma_0^2\approx3$ and $M^2_{\perp}(0)/\sigma_0^2\approx1$, and for the fluctuation initial condition $M^2(0)/Q^2\approx0.43$ for all coupling values we consider. All results are checked also for possible infrared cutoff effects and found to be stable.

The computations are performed in momentum space on an isotropic grid. In order to be able to reach sufficiently large times, we adapt discretizations as a function of time. While the computations are done for different sets of discretization parameters to check convergence, results shown for fluctuation (field) initial conditions for times $Q t \le 1200$ ($\sigma_0 t \le 1200)$ are from computations on a spatial grid characterized by $N_s=540$ points with a grid spacing $Q a_s=0.4$ ($\sigma_0 a_s = 0.5$). Subsequently, we reduce the number of grid points to $N_s=135$ ($N_s=225$), thereby increasing the infrared cutoff. We checked that the spectra lie on top of each other in the common momentum range for overlapping times. In general, the temporal grid at early times has to be finer than at later times in order to resolve the initially faster processes such as instabilities as compared to the subsequent slow power-law behavior near nonthermal fixed points. For the displayed results, we changed from $Q a_t = 0.08$ ($\sigma_0 a_t = 0.15$) for early times to $Q a_t = 0.16$ ($\sigma_0 a_t = 0.30$) at $Q t = 160$ ($\sigma_0 t = 225$). For field initial conditions at stronger couplings $\lambda \gtrsim 1$, it turned out that it is sufficient to use $N_s = 225$ at all times, however, we employed smaller step sizes changing from $\sigma_0 t = 0.10$ to $0.20$ at $\sigma_0 t = 150$.

\section{\label{sec:FluctuationIC}Results Starting from Fluctuation Initial Conditions}

\subsection{\label{secDistributionAndDispersion}Distribution and Dispersion Relation}

We first consider the dynamics in the symmetric regime starting from the fluctuation initial condition (a) of Sec.~\ref{sec:initialConditions}. In order to discuss the nonequilibrium time evolution, it is convenient to extract a time-dependent particle number distribution $f(t,\pabs)$, which can be defined in terms of the equal-time statistical two-point function~\coloreqref{eq:defF} in spatial Fourier space as~\colorcite{Berges:2001fi}
\begin{equation}
\frac{f(t,\pabs)+1/2}{\omega(t,\pabs)} =  \lambda^{-1} F(t,t,\pabs)\, .
\label{eq:distributionTwoPointFunction}
\end{equation}
At time $t=0$, this definition coincides with~\coloreqref{eq:initialConditionsF} for the initial dispersion~\coloreqref{eq:initialDispersion}. In practice, we extract a time-dependent distribution function and dispersion relation in accordance with~\coloreqref{eq:distributionTwoPointFunction} by employing
\begin{align}
\!\!\!f(t,\pabs)+\frac{1}{2} 	&= \lambda^{-1}\! \left. \sqrt{F(t,t^\prime,\pabs)\, \partial_{t}\partial_{t^\prime}F(t,t^\prime,\pabs)} \, \right|_{t=t^\prime}\, ,	\label{eq:distributionFunction}	\\
\omega(t,\pabs) 			&= \left. \sqrt{\partial_{t}\partial_{t^\prime}F(t,t^\prime,\pabs)/F(t,t^\prime,\pabs)} \, \right|_{t=t^\prime}\, .					\label{eq:dispersionRelation}
\end{align}		 
In general, there is, of course, no unique definition of a mode particle number in the interacting theory. Nevertheless, we will use the notion of a distribution function to illustrate our results, remembering that one can always think of the well-defined correlation functions on which our definitions are based.

\subsection{\label{sec:NonthermFixedPointsSymmetric}Nonthermal Fixed Points}

\begin{figure}
\subfloat{%
\includegraphics*{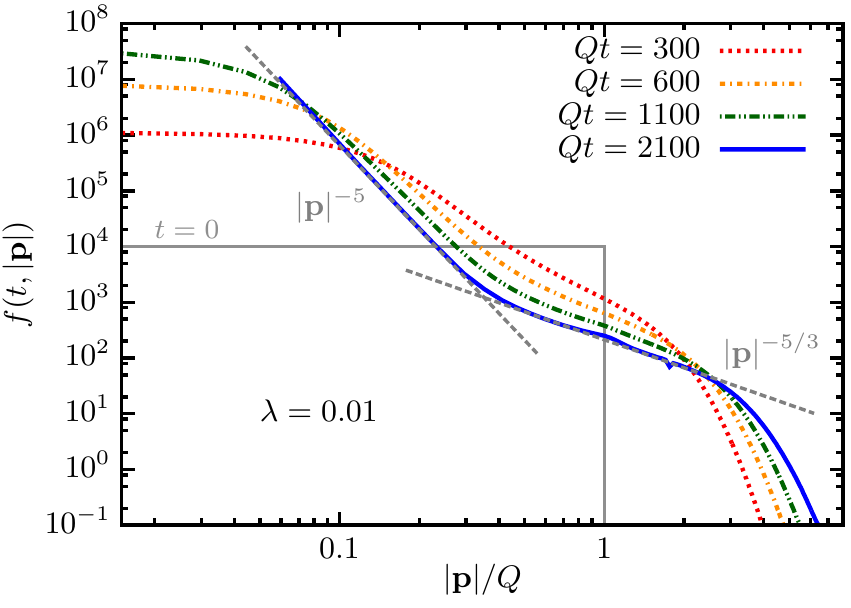}%
}\vspace{-3pt}
\subfloat{%
\includegraphics*{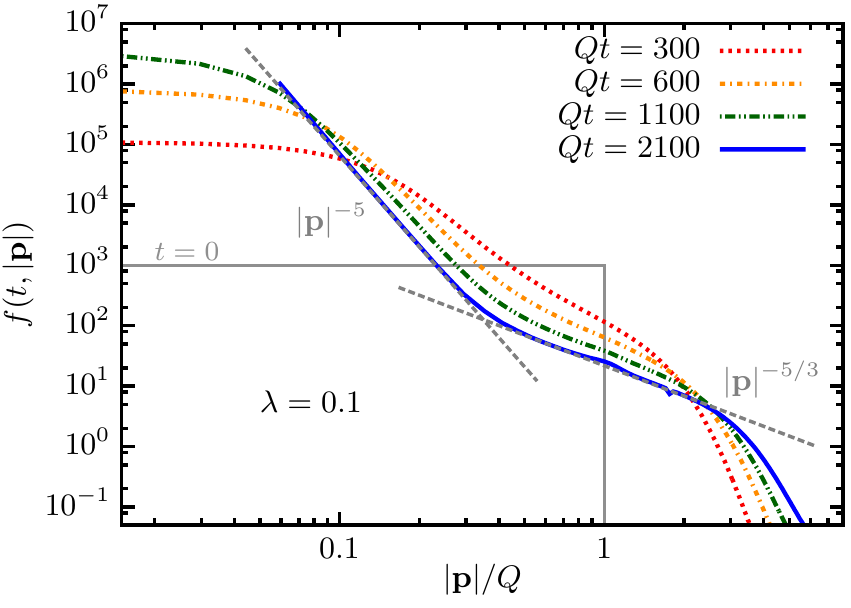}%
}\vspace{-3pt}
\subfloat{%
\includegraphics*{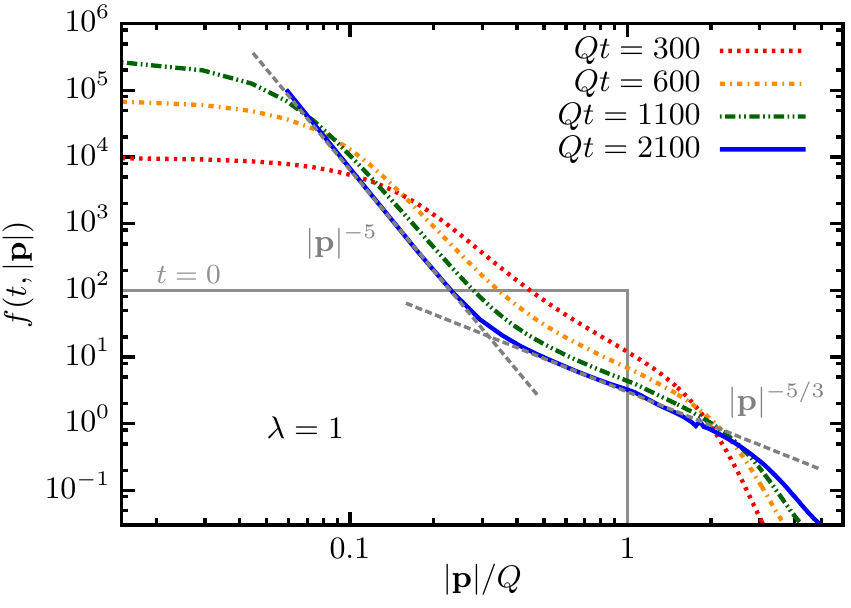}%
}
\caption{\label{fig:f_l001+01+1_f-p}Particle number distribution for fluctuation initial conditions with $\lambda=0.01, 0.1, 1$ (\textit{top} to \textit{bottom}) at different times including the initial time (gray). The approximate power laws with exponents $\kappa_\mathrm{N}=5$ and $\kappa_\mathrm{E}=5/3$ characterize the dual cascade  in the respective momentum ranges (dashed gray lines).\vspace{-1.3cm}}
\end{figure}

The distribution function for different values of the coupling $\lambda=0.01,\ 0.1\text{ and }1$ is presented in Fig.~\ref{fig:f_l001+01+1_f-p} at different times, which show the evolution starting from the displayed initial overoccupied state. One observes the emergence of two distinct regimes for small ($\pabs\lesssim0.3\,Q$) and for larger momenta ($\pabs\gtrsim0.3\,Q$). The evolution is self-similar~\colorcite{Orioli:2015dxa}, and the approximate power-law behavior corresponds to an inverse particle cascade towards the infrared~\colorcite{Berges:2008wm, Berges:2008sr} and a direct energy cascade towards larger momenta~\colorcite{Micha:2004bv}. This is the first time that the dual cascade has been observed to emerge in a quantum field theoretical calculation without relying on the classical-statistical approximation.

In fact, the scaling exponent for the direct energy cascade at higher momenta is found to be well approximated by
\begin{equation}
\kappa_\mathrm{E}=5/3 \, ,
\label{eq:kappa_energyCascade_fluctuationIC}
\end{equation}
which is the value predicted by analytical considerations in the framework of kinetic theory with two-to-two scattering in the absence of a macroscopic field expectation value ($\phi = 0$)~\colorcite{Micha:2004bv}.  The energy cascade exponent~\coloreqref{eq:kappa_energyCascade_fluctuationIC} has not been observed before from classical-statistical simulations. The latter see also a direct energy cascade, however, with an expected exponent $3/2$ characterizing the scattering of a particle off a macroscopic field mode~\colorcite{Micha:2004bv} (cf.~also Sec.~\ref{sec:FieldIC} below).

The possibility to observe~\coloreqref{eq:kappa_energyCascade_fluctuationIC} is due to the fact that we are able to preserve the $\mathrm{O}(N)$ symmetry at all times by setting the initial field $\phi(t=0)$ as well as its time derivative to zero. In contrast, classical-statistical simulations always start with nonzero values for the field or its time derivative. A vanishing macroscopic field average can then only be obtained by sampling over many individual runs, however, the initial bias cannot be set to zero identically with that procedure. The situation is similar to the well-known phenomenon of spontaneous symmetry breaking in thermal equilibrium, where a small bias can lead to the appearance of a macroscopic field value. 

\begin{figure}
\includegraphics*{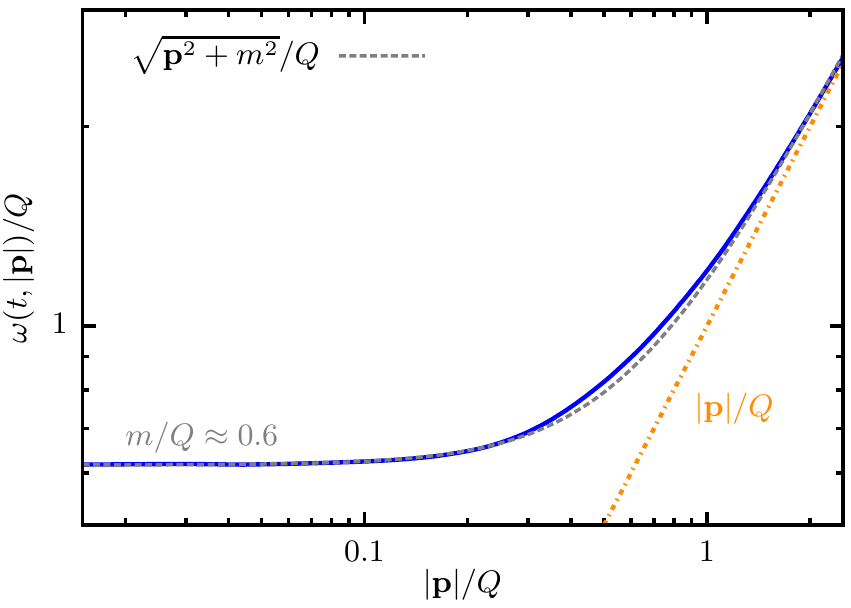}%
\caption{\label{fig:f_l001_disprel}Dispersion relation for fluctuation initial conditions with $\lambda=0.01$ at $Q t=1200$ exhibiting an effective mass gap $m$. A fit to the quasiparticle dispersion relation $\sqrt{\pvec^2+m^2}$ is performed in good agreement with the numerical data for $m/Q\approx0.6$. Additionally, a linear dispersion relation, \mbox{$\omega=\pabs$}, is shown for comparison.}
\end{figure}

{The inverse particle cascade at low momenta is well approximated by a power law with exponent
\begin{equation}
\kappa_\mathrm{N}=5
\end{equation}
as displayed by the gray dashed line in all three panels of Fig.~\ref{fig:f_l001+01+1_f-p}. The value of this exponent is characteristic for a nonrelativistic inverse particle cascade towards the infrared~\colorcite{Scheppach:2009wu, Orioli:2015dxa}. The appearance of the nonrelativistic infrared regime within the relativistic field theory is a consequence  of the presence of an effective mass gap, which has been pointed out in~\colorcite{Orioli:2015dxa}. Indeed, we see in Fig.~\ref{fig:f_l001_disprel} that the dispersion relation~\coloreqref{eq:dispersionRelation} is approximately constant in the momentum range where we observe the power law in the infrared. We also show a linear dispersion relation as expected for relativistic momenta. In addition, we fit a quasiparticle dispersion relation $\sqrt{\pvec^2+m^2}$ with the in-medium mass $m$ to the numerical results. We find that\unskip\parfillskip0pt\par}%
{~\vspace{3ex}\\\noindent}the dispersion relation fit remains almost constant over time during the turbulent stage of the evolution and exhibits an effective mass gap of $m\approx 0.6\,Q$. This plot is obtained from computations with $\lambda=0.01$ and shown for $Q t=1200$, but to very good accuracy the same graph and effective mass are found for the entire range of couplings, $\lambda=\numrange{0.01}{10}$, and not too early times.

\begin{figure}[b]
\begin{minipage}{\textwidth}
\subfloat[\label{fig:f_l10_f-p_f}Particle number distribution.]{
\includegraphics*{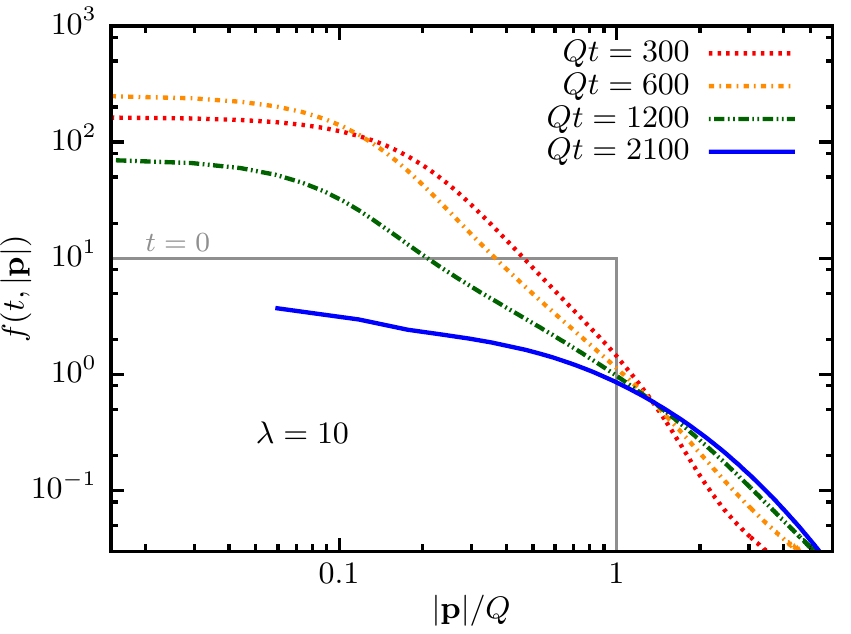}%
}\hfill
\subfloat[\label{fig:f_l10_f-p_ln}Inverse slope parameter $\ln(1+1/f)$.]{
\includegraphics*{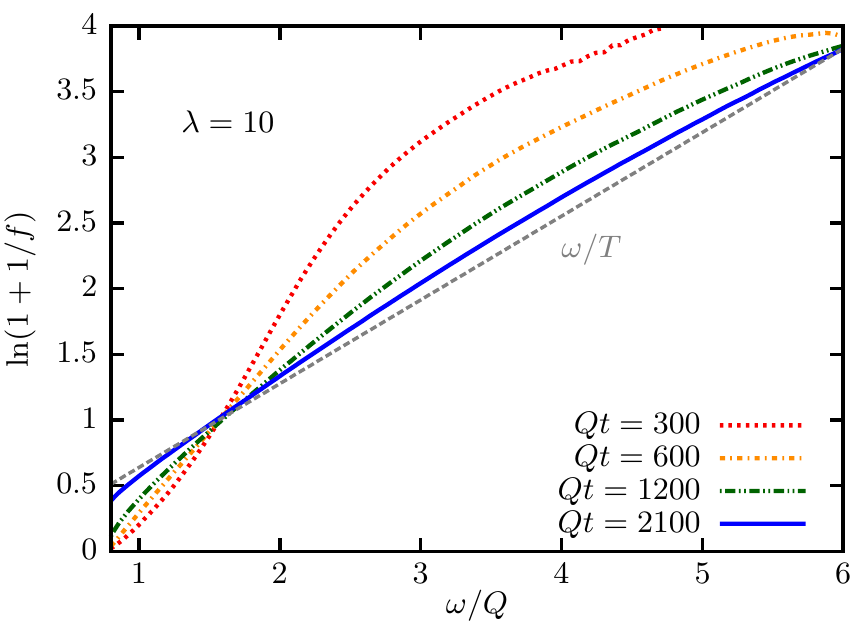}%
}
\caption{\label{fig:f_l10_f-p}Particle number distribution~(\textit{left}) and inverse slope parameter $\ln(1+1/f)$~(\textit{right}) for fluctuation initial conditions with $\lambda=10$ at different times including the initial time (solid, gray). We observe a power-law behavior for intermediate times and the approach to thermal equilibrium for $Q t\gtrsim 600$. For comparison, the dashed gray line in~\protect\subref{fig:f_l10_f-p_ln} corresponds to a Bose-Einstein distribution with temperature $T\approx1.6\,Q$ and chemical potential $\mu=0$.}
\end{minipage}
\end{figure}

Most remarkably, our results demonstrate that the system with coupling $\lambda=1$ still shows the same dual cascade picture as for weak couplings. From the lowest panel of Fig.~\ref{fig:f_l001+01+1_f-p}, we see that characteristic occupation numbers for the direct energy cascade have reached $\mathcal{O}(1)$, where quantum corrections are expected to start to become relevant. This demonstrates that even for sizable couplings of order one the perturbatively expected scaling properties for the energy cascade still hold in this case. Moreover, we conclude that the  nonperturbative infrared scaling properties that have been observed before from classical-statistical simulations~\colorcite{Berges:2008wm, Nowak:2011sk, Orioli:2015dxa} extend well beyond the weak-coupling limit.

\vskip4.5pt
We find for even stronger couplings important corrections to the above dual cascade picture, but the phenomenon of a significant transient increase of infrared occupancies turns out to be remarkably robust. This is demonstrated in Fig.~\ref{fig:f_l10_f-p_f} for the larger coupling $\lambda=10$. One observes that for $Q t\lesssim 600$ the fluctuations in the infrared grow by more than an order of magnitude. This is much less compared to the growth of fluctuations for smaller interaction strengths, which exhibit a growth by three orders of magnitude in the same period of time. Remarkably, the results still suggest an approximate transient power-law behavior during this time. For later times, the infrared occupancies decrease significantly. Moreover, the occupation numbers become of order unity at the characteristic scale $Q$ and the distribution function approaches a Bose-Einstein distribution with vanishing chemical potential, $f_\mathrm{BE}(\omega) = \left(\mathrm{e}^{\omega/T}-1\right)^{-1}$. Figure~\ref{fig:f_l10_f-p_ln} indicates this with the dotted line representing a thermal distribution with temperature $T\approx1.6\,Q$. The fact that we are able to explicitly demonstrate the approach to quantum thermal equilibrium in this case has to do with the fact that it happens faster for stronger couplings, while it becomes numerically too expensive for small couplings to go to much later times.

\begin{figure}
\centering
\includegraphics*{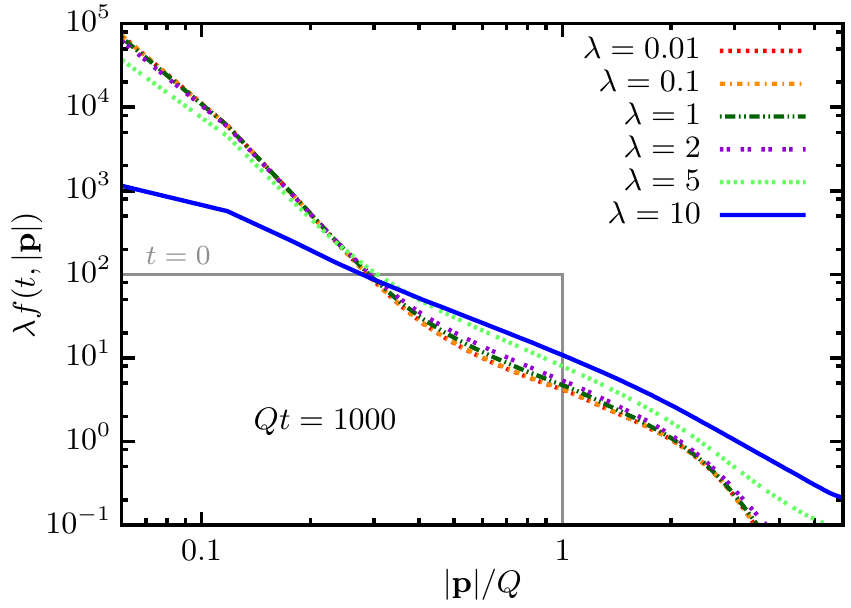}%
\caption{\label{fig:f_reparam_f-p}Rescaled particle number distribution for fluctuation initial conditions at fixed time for different values of the coupling $\lambda$. In the range $\lambda\lesssim 2$, one observes that all curves collapse onto a single curve to good accuracy. In contrast, for stronger couplings sizable deviations occur.}
\end{figure}

\vskip4.5pt
{The above observation, that even for couplings of order one we find transient universal scaling behavior, can\unskip\parfillskip 0pt \par}{\newpage\noindent}be well illustrated by plotting the product $\lambda f(t,\pvec)$. This product should be independent of the value of $\lambda$ in the scaling regime.  Figure~\ref{fig:f_reparam_f-p}%
shows the quantity on spatial grids with $N_s=135$ for $Q t = 1000$. Up to coupling constants of $\lambda\lesssim 2$, the rescaled distributions lie on top of each other to good accuracy. Differences between the rescaled distribution functions do not become sizable until the coupling has increased to about $\lambda\gtrsim 5$. This demonstrates the existence and range of attraction  of the nonthermal fixed point for a wide range of couplings well beyond the weak-coupling limit.

\subsection{\label{sec:NonperturbativeInfraredSymmetric}Nonperturbative Infrared Regime}
A quantity that illustrates the nonperturbative nature of the nonthermal infrared scaling regime is given by the sum of all ring diagrams entering as the main building block in the $1/N$ expansion at NLO as discussed in Sec.~\ref{sec:Theory} and App.~\ref{app:2piEoM}.\footnote{The relation of the ring sum at NLO in the large-$N$ expansion to an effective vertex resummation is explained in~\colorcite{Aarts:2002dj, Berges:2010ez, Orioli:2015dxa}.} The summation of all ring diagrams is encoded in the self-consistent equations~\colorcite{Berges:2001fi}
\begin{align}
I_F(t,t^\prime,\pvec) 	&= \Pi_F(t,t^\prime,\pvec) - \int_{t_0}^t \!\!\d t^{\prime\prime}\: I_\rho(t,t^{\prime\prime},\pvec) \Pi_F(t^{\prime\prime},t^\prime,\pvec)				\nonumber 		\\
					&\phantom{=\ }+ \int_{t_0}^{t^\prime} \!\!\d t^{\prime\prime}\: I_F(t,t^{\prime\prime},\pvec) \Pi_\rho(t^{\prime\prime},t^\prime,\pvec)				\label{eq:IF}\, ,	\\
I_\rho(t,t^\prime,\pvec)&=\Pi_\rho(t,t^\prime,\pvec) - \int_{t^\prime}^{t} \!\!\d t^{\prime\prime}\: I_\rho(t,t^{\prime\prime},\pvec) \Pi_\rho(t^{\prime\prime},t^\prime,\pvec)\, ,	\label{eq:Irho}
\end{align}
which appear in the nonequilibrium time-evolution equations of App.~\ref{app:2piEoM}, with $\Pi_F$ denoting the real part of the one-loop self-energy given by~\coloreqref{eq:oneLoopSelfEnergy} and the corresponding imaginary part $\Pi_\rho(t,t^\prime,\pvec) = F_{ab}(t,t^\prime,\pvec) \rho_{ab}(t,t^\prime,\pvec)/(3N)$. By iteration, the above equations can be seen to generate the ring diagrams, which are contained in $I_F$ and $I_\rho$, to infinite loop order. It is instructive to compare the one-loop self-energy $\Pi_F(t,t,\pvec)$ at equal times to the one including all NLO large-$N$ corrections as given by $I_F(t,t,\pvec)$. More precisely, we consider the ratio
\begin{equation}
\frac{I_F(t,t,\pvec)}{\Pi_F(t,t,\pvec)} = 1 - \frac{\int_{t_0}^t \!\d t^{\prime\prime} \left( I_\rho \Pi_F - I_F \Pi_\rho\right)}{\Pi_F(t,t,\pvec)} \, ,
\label{eq:ratio}
\end{equation}
where the r.h.s.\ follows from~\coloreqref{eq:IF} evaluated at equal times, $t=t^\prime$. In a perturbative regime, we expect $I_F$ to be dominated by the lowest one-loop contribution $\Pi_F$ such that the ratio~\coloreqref{eq:ratio} is about one. 

\begin{figure}[ht!]\vspace{1.1pt}
\subfloat{%
\includegraphics*{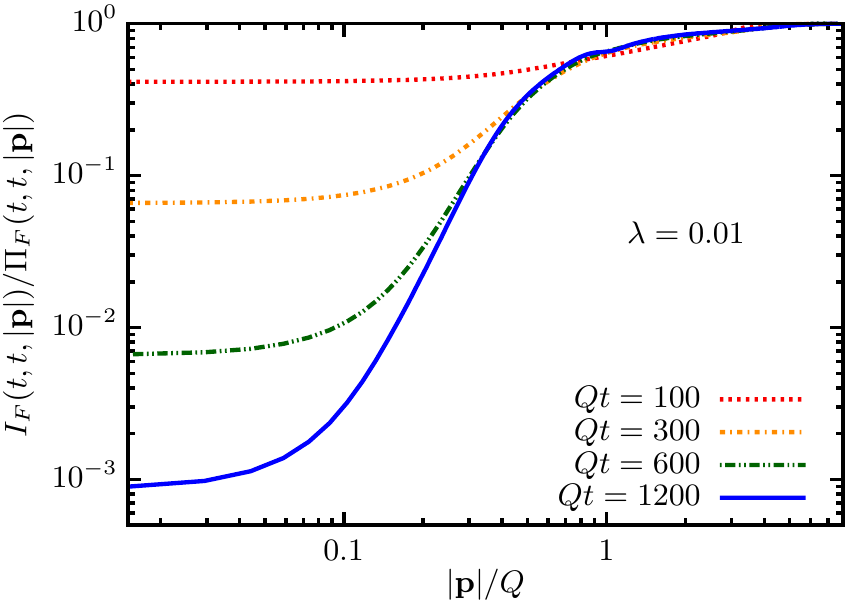}%
}\vspace{-1pt}\vspace{-5pt}
\subfloat{%
\includegraphics*{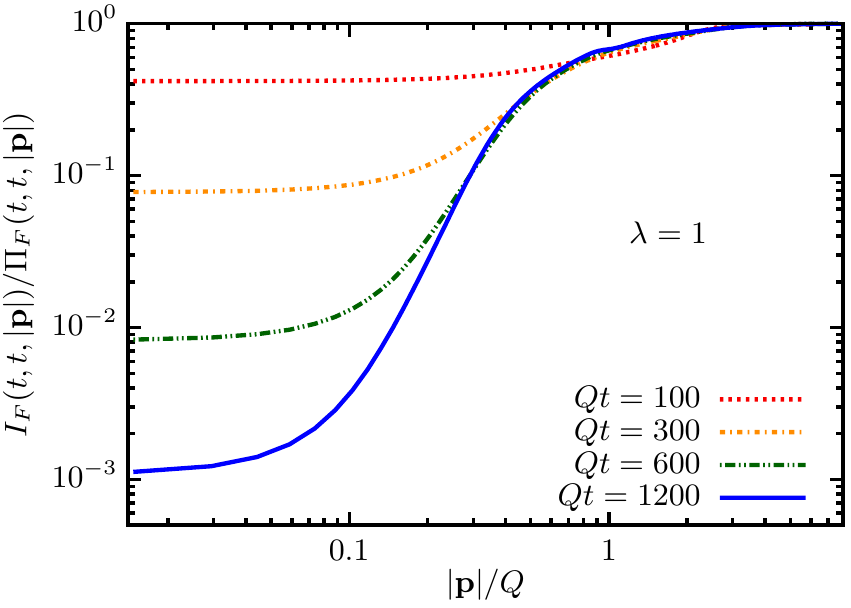}%
}\vspace{-1pt}\vspace{-5pt}
\subfloat{%
\includegraphics*{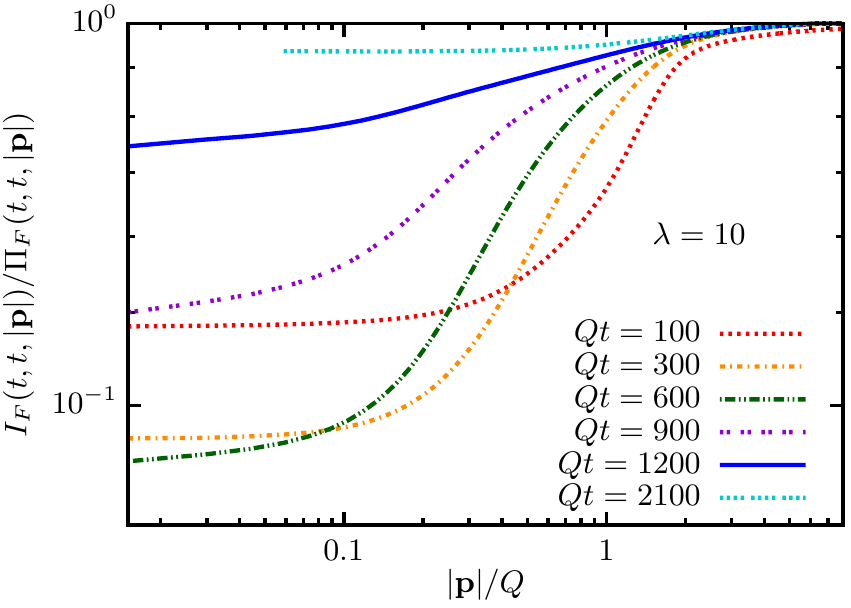}%
}\vspace{-5pt}
\caption{\label{fig:f_l001+1+10_ratio-p}The ratio~\coloreqref{eq:ratio} for fluctuation initial conditions with $\lambda=0.01, 1, 10$ (\textit{top} to \textit{bottom}). In a perturbative regime, $I_F$~is expected to be dominated by the lowest one-loop contribution~$\Pi_F$ such that the ratio~\coloreqref{eq:ratio} is about one. While at high momenta the ratio approaches unity, in the infrared regime one observes very strong deviations from one for times where the transient inverse particle cascade occurs.\vspace{-1ex}}
\end{figure}

\vskip4pt
From the top panel of Fig.~\ref{fig:f_l001+1+10_ratio-p} for $\lambda=0.01$, one observes that the ratio~\coloreqref{eq:ratio} decreases over several orders of magnitude in the low momentum regime during those times where the transient inverse particle cascade occurs according to the results of Sec.~\ref{sec:NonthermFixedPointsSymmetric}. Around the inflection point of the $I_F/\Pi_F$ curve on a double logarithmic scale, we find an approximate slope corresponding to a drop in momenta $\sim \pabs^{3.4}$ at $Q t=1200$. At large momenta, the ratio approaches unity as anticipated, but still differs sizably from one in the momentum range of the energy cascade.

We have verified that the values found for $\lambda=0.1$ agree well with those obtained for $\lambda=0.01$, which holds to good accuracy even for $\lambda=1$ as demonstrated with the middle panel of Fig.~\ref{fig:f_l001+1+10_ratio-p}. For $\lambda=10$, the bottom panel of Fig.~\ref{fig:f_l001+1+10_ratio-p} shows significant deviations from unity in the infrared at intermediate times, while  the ratio approaches unity again for later times where the $\lambda=10$ results already thermalize. This reflects the fact that occupation numbers first increase for small momenta as a consequence of the inverse particle cascade and then approach order unity on their way to thermal equilibrium.

\subsection{\label{sec:classStatSymmetric}Classical-Statistical Field Theory Limit}
As discussed in Sec.~\ref{sec:Theory} (cf.~also App.~\ref{app:2piEoM}), we can study the corresponding classical-statistical field theory in the same framework of the $1/N$ expansion to NLO following along the lines of~\colorcite{Aarts:2001yn}. In~\colorcite{Berges:2008wm}, the agreement of the quantum dynamics%
with results from classical-statistical simulations during parametric resonance at early times for the same model was shown to agree well for weak couplings. A similar study~\colorcite{Arrizabalaga:2004iw}, employing the tachyonic instability, also reports a matching of both methods.

The range of validity of the classical-statistical approach was also studied extensively in~\colorcite{Berges:2013lsa} for a massless one-component scalar field theory. In particular, Ref.~\colorcite{Berges:2013lsa} reported that if the classical-statistical evolutions are started from quantum initial conditions, the subsequent dynamics can lead to the decay of the initial quantum-half if the couplings are too strong. Here, we add to these findings by a direct comparison between the quantum and classical evolutions, i.e.~by solving the evolution equations at NLO in the $1/N$ expansion both for the classical-statistical theory and for the quantum theory starting from the same fluctuation initial conditions. 

\begin{figure}
\includegraphics*{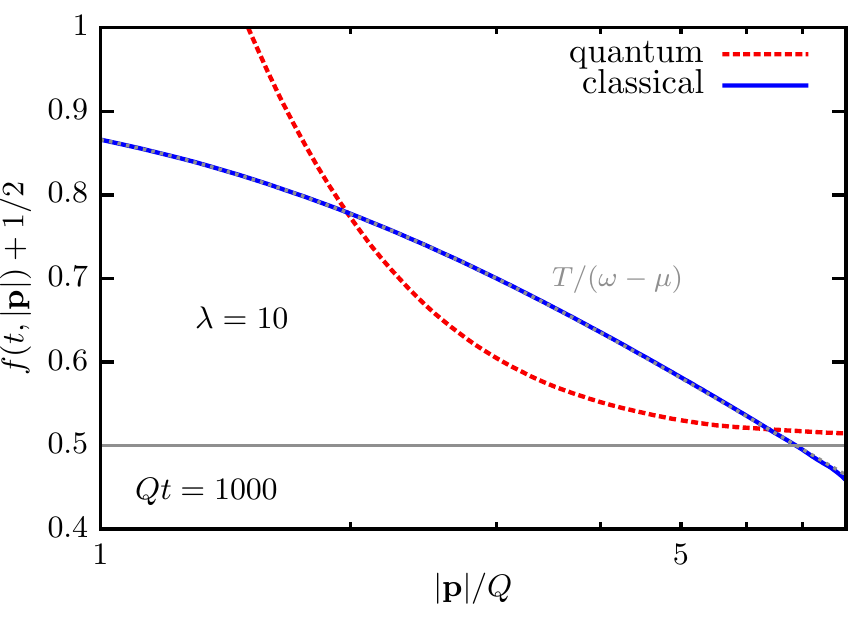}%
\caption{\label{fig:f_l10_comp_f-p}The sum $f(t,\pabs)+1/2$ for $\lambda=10$ at time $Qt = 1000$ for the quantum (dashed line) and the classical (solid line) evolution for fluctuation initial conditions. In addition, we show the classical thermal function $T/(\omega(\pvec)-\mu)$ (gray, dotted) for comparison, with parameters as described in the main text. One observes that, in contrast to the quantum evolution, the classical results for $f(t,\pabs)+1/2$ drop below $1/2$ at high momenta, showing the decay of the initial quantum-half.}
\end{figure}

\vskip4pt
Figure~\ref{fig:f_l10_comp_f-p}%
shows our results for $\lambda=10$ at time $Qt = 1000$ for the quantum (dashed line) and the classical (solid line) evolution. At this time, we find that the sum $f(t,\pabs)+1/2$ as obtained from the classical evolution can  already be described by a classical thermal distribution $T/(\omega-\mu)$ with temperature $T$, dispersion relation $\omega=\sqrt{\pvec^2+m^2}$ and chemical potential $\mu$. Since the classical thermal distribution leads to the Rayleigh-Jeans divergence, all classical thermal results depend on the employed ultraviolet cutoff $\Lambda_\mathrm{UV}= 7.9Q$. For this cutoff, we find $T\approx6.7Q$ and $\mu\approx-6.5Q$ at $Q t=1000$ with an in-medium mass $m\approx0.7Q$, where the latter was obtained from the numerically computed dispersion relation as in Sec.~\ref{sec:NonthermFixedPointsSymmetric}.	

\begin{figure*}
\vspace{-4mm}
\subfloat{
\includegraphics*{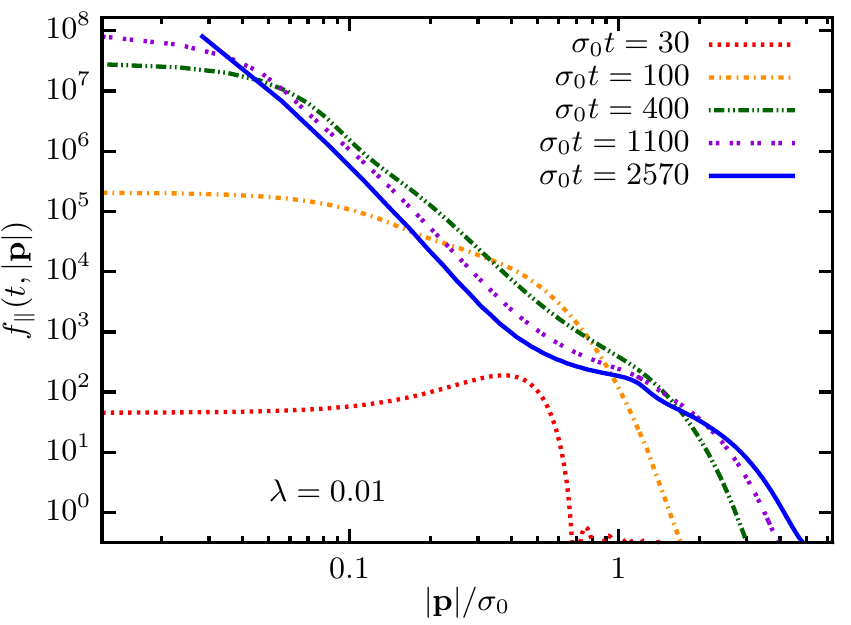}%
}\hfill
\subfloat{
\includegraphics*{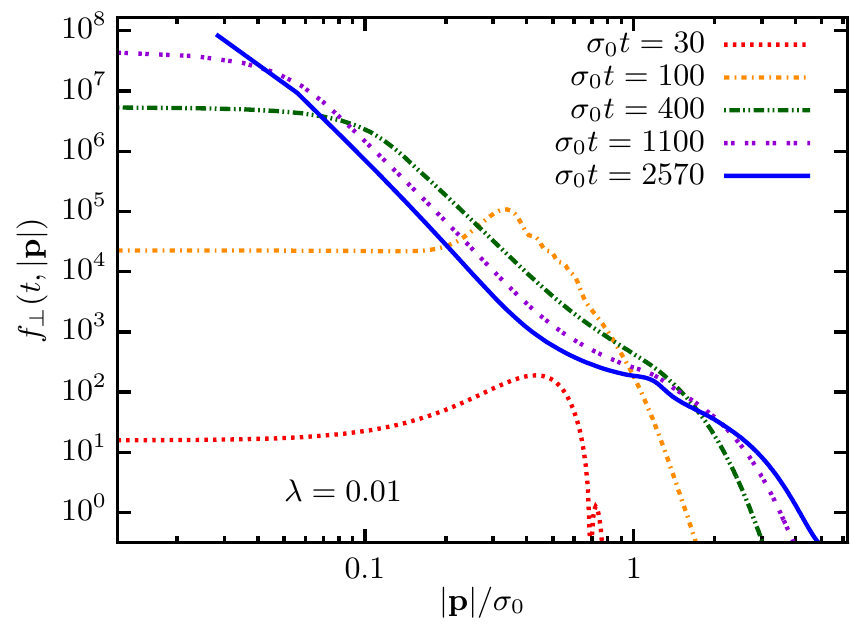}%
}\vspace{-1ex}
\caption{\label{fig:phi_l001_f-p}Longitudinal ($\parallel$, \textit{left}) and transverse ($\perp$, \textit{right}) occupation number distributions for macroscopic field initial conditions with $\lambda=0.01$ at different times showing the emergence of a dual cascade picture.}
\end{figure*}

By comparison to the horizontal line in Fig.~\ref{fig:f_l10_comp_f-p}, we see that for the classical evolution the sum $f(t,\pabs)+1/2$ drops below $1/2$ for high momenta. This shows the decay of the initial vacuum quantum-half and, hence, is outside the range of validity of the classical-statistical approximation to describe the quantum dynamics in this case. In contrast, in the quantum theory, $f(t,\pabs)$ never reaches negative values and is seen to approach a Bose-Einstein distribution at late times. Of course, classical dynamics cannot describe the approach to quantum thermal equilibrium and, for given finite coupling and ultraviolet regularization, the classical description is only valid for a restricted time. The smaller the coupling, the later the breakdown of the classical-statistical approximation occurs, which is also what we observe.\footnote{In general, for a self-interacting boson field $\varphi$ with coupling $\lambda$, the mapping between the quantum and the classical description involves two steps~\colorcite{Son:1996uv, Khlebnikov:1996mc}: First, one separates the quantum field $\varphi=\phi + \delta \varphi$ into a (large) coherent part with expectation value $\langle \varphi \rangle = \phi \sim Q/\sqrt{\lambda}$ and a (small) fluctuation with $\langle \delta \varphi^2 \rangle \sim Q^2$ for some given momentum scale $Q$ and coupling parameter $\lambda \ll 1$. The early-time dynamics can then be linearized in the fluctuations as long as $\langle \delta \varphi^2 \rangle \ll Q^2/\lambda$ and solved. Albeit small initially, the fluctuations can grow with time because of nonequilibrium instabilities. Once the fluctuations become larger, one uses the outcome of the linearized early-time evolution as input for a subsequent fully nonlinear classical-statistical simulation. The whole procedure has a well-defined continuum limit for sufficiently weak couplings (see also~\colorcite{Aarts:1997kp}). Instead of this two-step procedure, a simplified description is often employed: the fully nonlinear classical-statistical simulation is considered right from the beginning as a shortcut procedure. In this case, for strong-field initial conditions the description is only valid as a low-energy effective theory for a finite time depending on the employed coupling and ultraviolet cutoff scale~\colorcite{Epelbaum:2014yja}.}

\section{\label{sec:FieldIC}Results Starting from Macroscopic Field Initial Conditions}

\subsection{\label{sec:NonthermalFixedPointsBroken}Nonthermal Fixed Points} 
In addition to the fluctuation initial conditions with vanishing macroscopic field discussed in the previous section, we also investigate the time evolution starting from a macroscopic field with only vacuum fluctuations as described in Sec.~\ref{sec:initialConditions}. We can see from Fig.~\ref{fig:phi_l001_f-p} for $\lambda = 0.01$ that the system builds up highly occupied modes in the infrared very quickly through the mechanism of parametric resonance with the initial characteristic resonance peak at $\pabs\sim\sigma_0/\sqrt{2}$~\colorcite{Berges:2002cz}. One observes that the early-time evolution of the longitudinal and transverse degrees of freedom differs, but they become almost indistinguishable from each other at later times. This is due to the fact that the macroscopic field decays with time.

Similar to what is found for fluctuation initial conditions, the inverse particle cascade in the infrared can be clearly observed from Fig.~\ref{fig:phi_l001_f-p}. A fit produces a value of $\kappa_\mathrm{N}\approx4.7$ for the associated stationary exponent at the available times, which is expected to increase further for later times. The dispersion relation of this system is approximately constant in the momentum range of the particle cascade, very similar to the relation obtained from the fluctuation initial conditions (cf.~Fig.~\ref{fig:f_l001_disprel}). Fitting a power law~$\sim\pabs^{-3/2}$ at higher momenta hints to a direct energy cascade~\colorcite{Micha:2004bv}, but the resonance peaks have not yet smoothed out at the latest times we consider.

\begin{figure*}
\vspace{-4mm}
\subfloat{
\includegraphics*{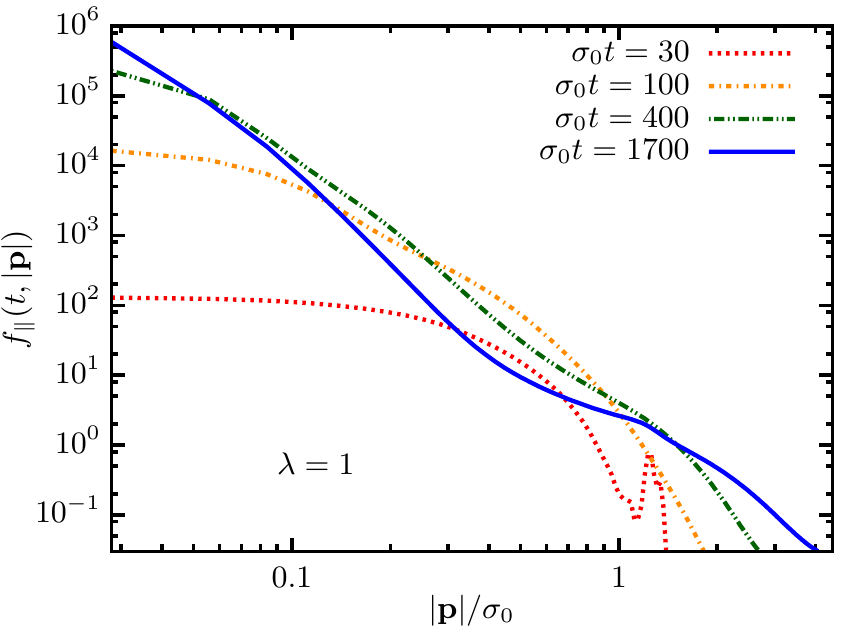}%
}\hfill
\subfloat{
\includegraphics*{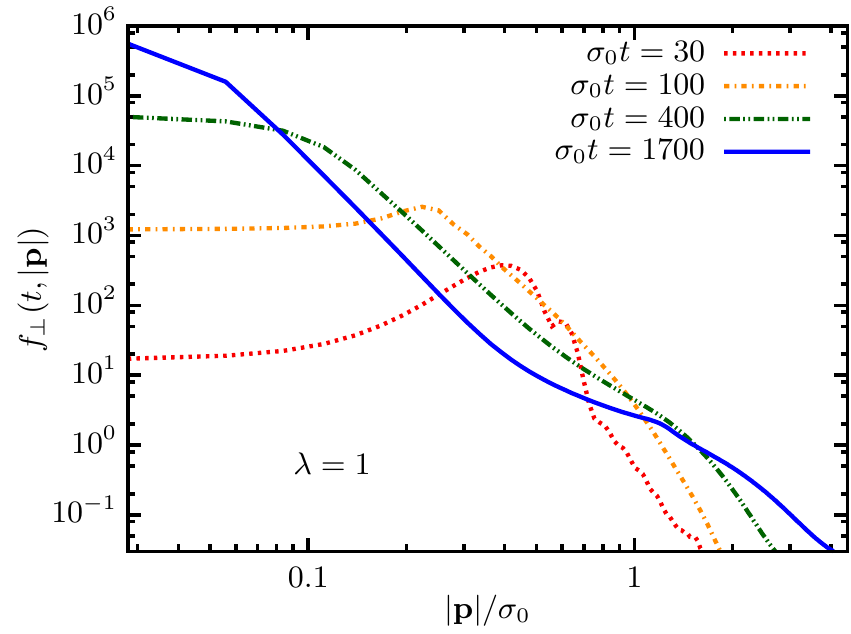}%
}\vspace{-1ex}
\caption{\label{fig:phi_l1_f-p}As in Fig.~\ref{fig:phi_l001_f-p}, but for $\lambda=1$, for which the system is still showing a dual cascade picture.\vspace{-3pt}}
\end{figure*}

The computations with an interaction strength of $\lambda=0.1$ yield practically the same distributions as just discussed for $\lambda=0.01$ up to a rescaling with the coupling as expected from the discussion for the symmetric case above. When increasing the coupling to $\lambda=1$, the overall picture as presented in Fig.~\ref{fig:phi_l1_f-p}%
remains very similar if compared to the weak-coupling results displayed in Fig.~\ref{fig:phi_l001_f-p}. We find that the particle cascade at low momenta can be fitted with power laws~$\sim\pabs^{-4.6}$ and~$\sim\pabs^{-4.8}$ for longitudinal and transverse modes, respectively. We note that they still increase to larger values over time. 

\begin{figure*}
\vspace{-4mm}
\subfloat{
\includegraphics*{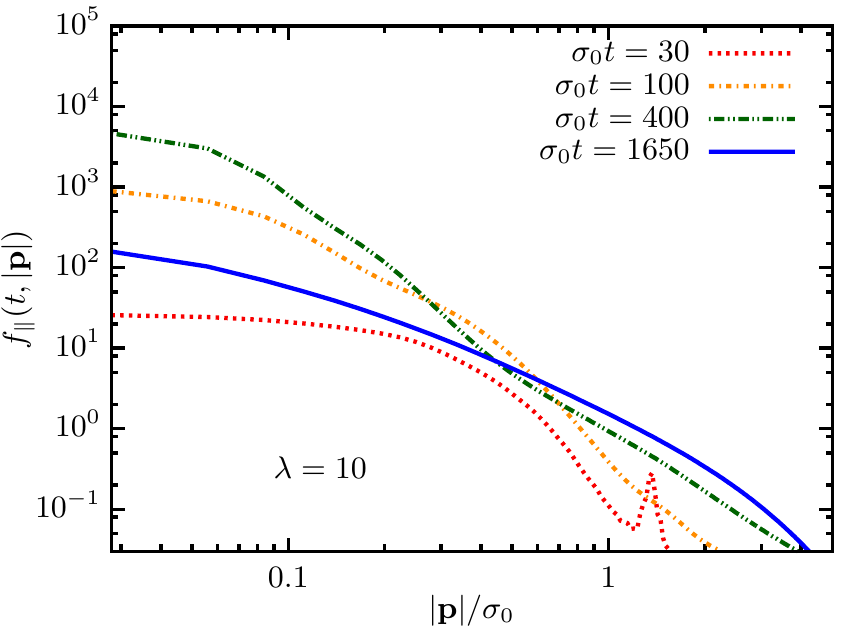}%
}\hfill
\subfloat{
\includegraphics*{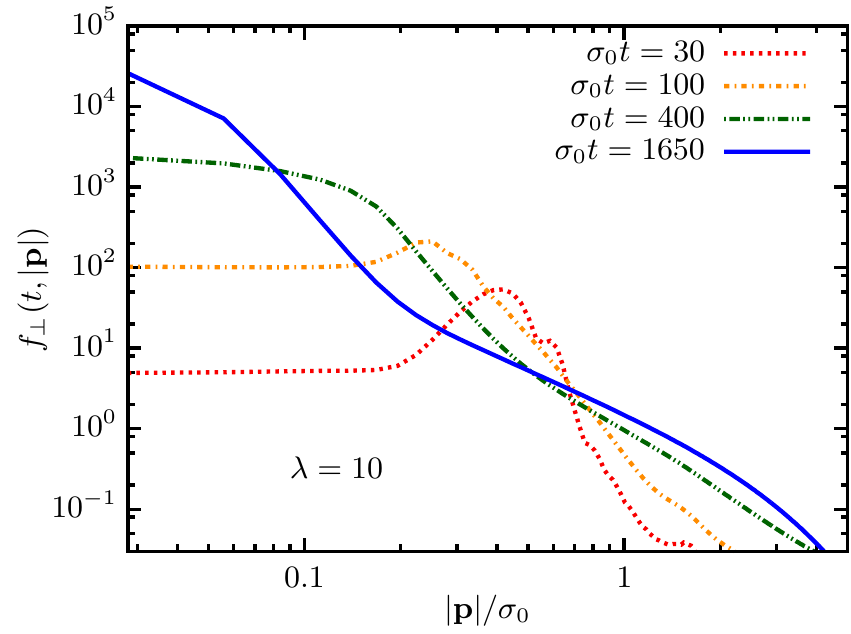}%
}\vspace{-1ex}
\caption{\label{fig:phi_l10_f-p}As in Fig.~\ref{fig:phi_l001_f-p}, but for $\lambda=10$, where strong deviations from the dual cascade picture occur. The longitudinal spectrum approaches thermal equilibrium rather quickly. At the same time, the transverse occupation number distribution exhibits a cascade-like behavior for small momenta, while the ultraviolet already shows signs of thermalization.}
\end{figure*}

A significantly modified picture emerges for the distribution functions for a coupling of $\lambda=10$, which we show in Fig.~\ref{fig:phi_l10_f-p}%
for both longitudinal and transverse modes. We observe an increase of the longitudinal occupation number in the infrared for $\sigma_0 t \lesssim 400$, with a subsequent decrease at later times. For large momenta, a quick approach to a thermal-like curve is observed, which is reminiscent of the situation described in Sec.~\ref{sec:NonthermFixedPointsSymmetric} for fluctuation initial conditions. At the same time, the transverse modes still exhibit a cascade-like behavior in the infrared, as can be seen from the evolution for times $\sigma_0 t\gtrsim 200$. Part of the curve in the infrared could be fitted with a power law~$\sim\pabs^{-4.5}$ at $\sigma_0 t=1650$. Both the scaling exponent and the distribution function in the deep infrared still grow at the latest time available and are far from equilibrium. Whereas the ultraviolet appears closer to thermal equilibrium for both the longitudinal and transverse degrees of freedom (they are approximately equal for $\pabs\gtrsim0.3\sigma_0$ at the latest time shown), they differ in the infrared by more than two orders of magnitude at $\sigma_0 t=1650$. 

\begin{figure*}
\vspace{-4mm}
\subfloat{
\includegraphics*{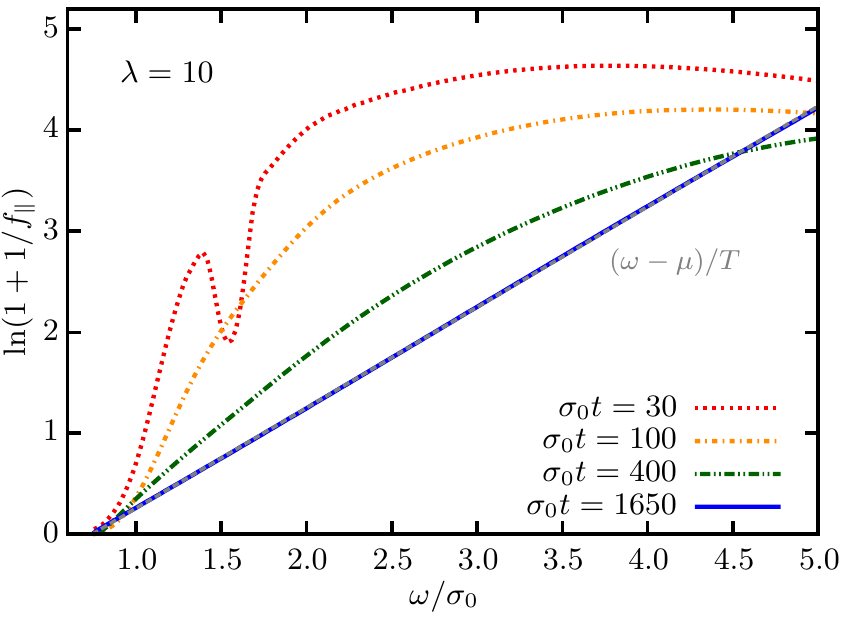}%
}\hfill
\subfloat{
\includegraphics*{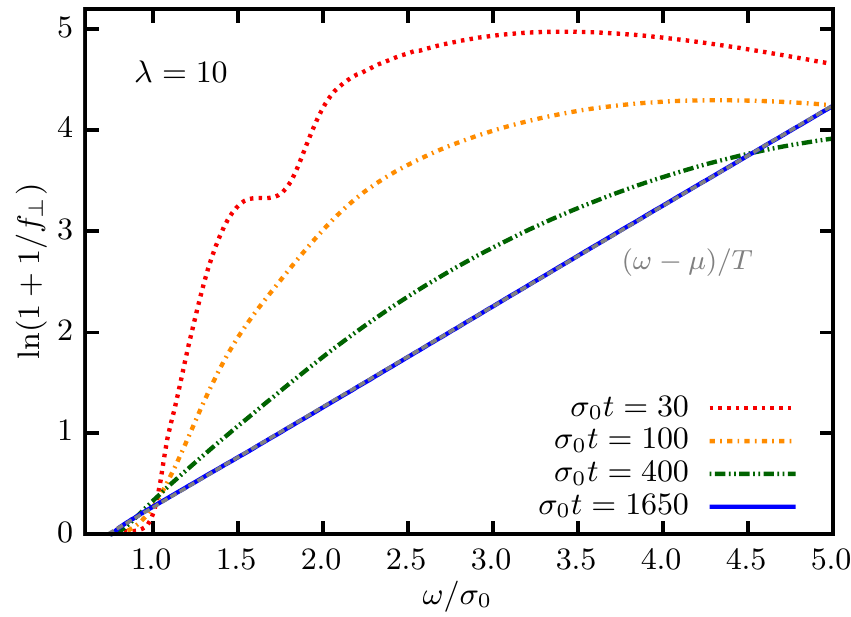}%
}\vspace{-1ex}
\caption{\label{fig:phi_l10_BEdistr_f-p}Longitudinal ($\parallel$, \textit{left}) and transverse ($\perp$, \textit{right}) inverse slope parameters $\ln(1+1/f_{\parallel,\perp})$ for macroscopic field initial conditions with $\lambda=10$ at different times. We observe the approach to thermal equilibrium. For comparison, the dashed line in each plot corresponds to a Bose-Einstein distribution with temperature $T\approx\sigma_0$ and chemical potential $\mu\approx0.7\sigma_0$.\vspace{-3pt}}
\end{figure*}

In Fig.~\ref{fig:phi_l10_BEdistr_f-p}, %
where the inverse slope parameter for both the longitudinal and transverse degrees of freedom is presented, we can explicitly see the approach to a Bose-Einstein distribution with temperature $T\approx \sigma_0$ and chemical potential $\mu\approx0.7\sigma_0$ at the latest available time ($\sigma_0 t = 1650$). For the real scalar field theory, we expect this distribution to still evolve until a vanishing chemical potential is reached~\colorcite{Berges:2001fi}. Moreover, the fast isotropization in field space is apparent as the curves are very similar for $\sigma_0 t = 100$ and already indistinguishable for $\sigma_0 t=400$.

\begin{figure*}
\vspace{-1ex}
\centering
\subfloat{
\includegraphics*{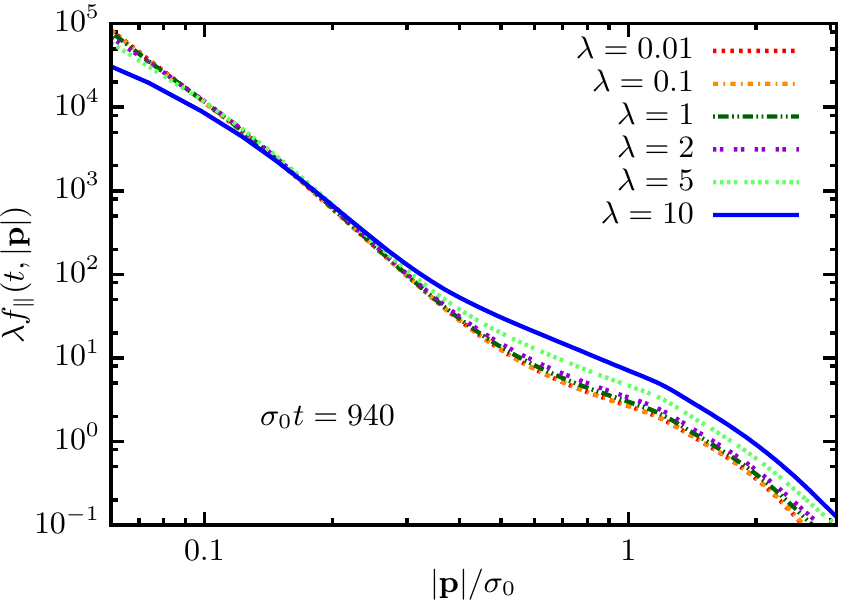}%
}\hfill
\subfloat{
\includegraphics*{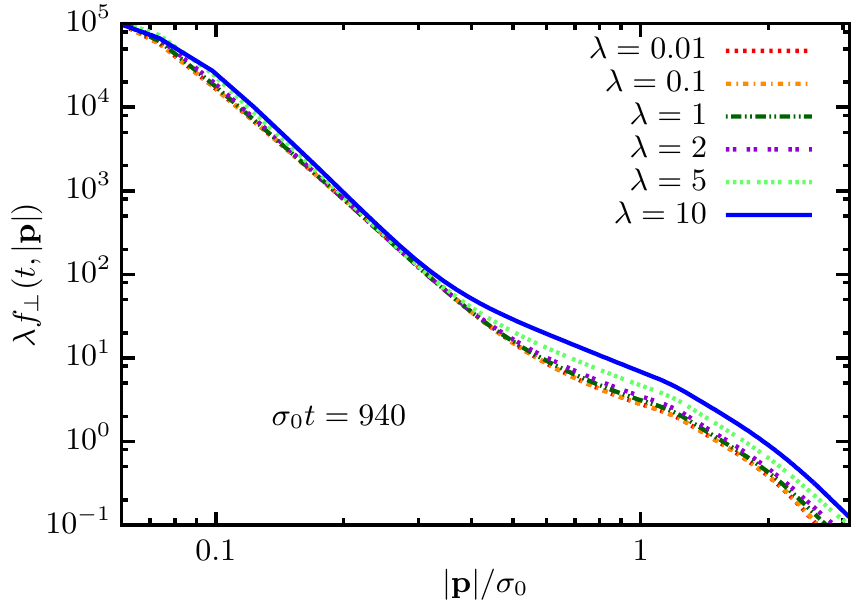}%
}\vspace{-1ex}
\caption{\label{fig:phi_reparam_f-p}Rescaled longitudinal ($\parallel$, \textit{left}) and transverse ($\perp$, \textit{right}) occupation number distributions for macroscopic field initial conditions at fixed time $\sigma_0 t=940$ for different values of the coupling $\lambda$.}
\end{figure*}

\vskip4pt
Finally, we present the rescaled distribution functions $\lambda f_{\parallel,\perp}(t,\pvec)$ for fixed times in the symmetry-broken regime in Fig.~\ref{fig:phi_reparam_f-p}, %
which are obtained through calculations where spatial grids with $N_s=128$, $\sigma_0 a_s=1.0$ and a temporal step size $\sigma_0 a_t=0.2$ are used. The rescaled spectra lie on top of each other to good accuracy well beyond the weak-coupling regime up to $\lambda\lesssim 2$. The deviations start to occur first in the energy cascade, whereas the particle cascade still exhibits good agreement. In the latter momentum range, differences between the rescaled distribution functions do not become sizable until the coupling has increased to $\lambda>5$. Moreover, the deviations for the transverse low-momentum modes are smaller if compared to the ones for the longitudinal direction, whereas they agree well for larger momenta.

\subsection{\label{sec:fieldDynamics}Macroscopic Field Dynamics}
	
\begin{figure}
\includegraphics*{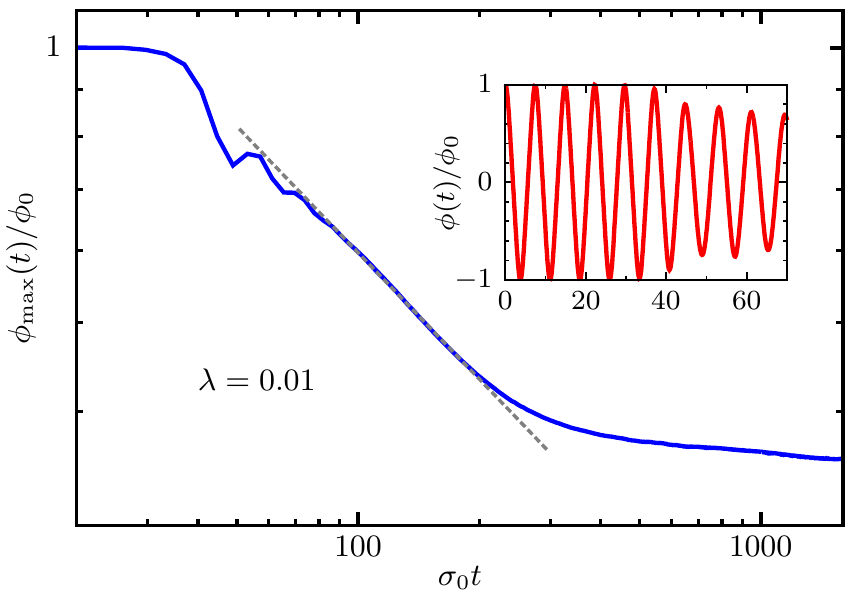}%
\vspace{-0.8ex}
\caption{\label{fig:phi_l001_phi-t}Double-logarithmic plot of the time evolution of the envelope  $\phi_\mathrm{max}(t)$ normalized to its initial value $\phi_0$ for $\lambda=0.01$. In the inset, the oscillating field is displayed. The field amplitude is~$\sim t^{-c}$ with $c\approx0.5$ at intermediate times and with a much smaller exponent of the order of a few percent during the turbulent stages of the evolution at later times.}
\end{figure}

In the following, we investigate the time evolution of the macroscopic field $\phi(t)$. Figure~\ref{fig:phi_l001_phi-t}%
shows the time evolution of the envelope of the macroscopic field amplitude, $\phi_\mathrm{max}$. The macroscopic field itself oscillates rapidly as can be inferred from the inset of the figure. The oscillation frequency retains its value over the computed period of time and is set by the same dynamically generated effective mass, $\omega_\phi\approx 0.7\sigma_0$, as inferred from the dispersion relation which has the same form as in the symmetric regime, cf.~Fig.~\ref{fig:f_l001_disprel}.

After the end of parametric resonance, there is an intermediate power-law decay of the field amplitude~$\sim t^{-c}$ with $c\approx0.5$. Comparing to the evolution of the distribution function, we observe that this decay happens during the approach to the self-similar turbulent regime. At about $\sigma_0 t\approx 200$, the field decay slows down and the amplitude may be described by a power law with a much smaller exponent. Moreover, the isotropization of the system in field space can be attributed to the decay of the macroscopic field as it becomes less important and leads to a decrease in the difference between the longitudinal and transverse degrees of freedom.

\begin{figure}
\includegraphics*{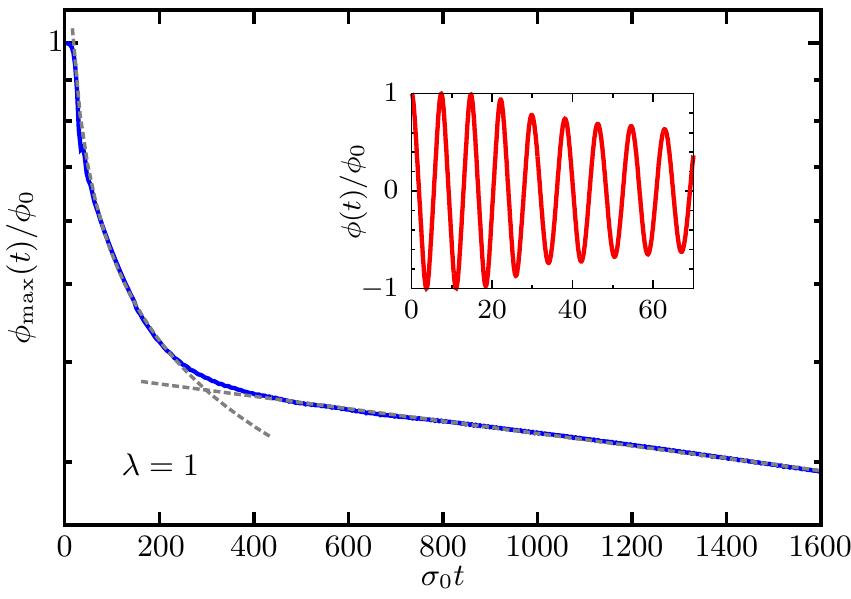}%
\vspace{-0.9ex}
\caption{\label{fig:phi_l1_phi-t}For $\lambda=1$, the power law of the field amplitude $\phi_\mathrm{max}(t)$ is~$\sim t^{-c}$ with $c\approx0.4$ for intermediate times (note the log-lin plot). At later times, an exponential damping with a small rate $\sim 10^{-4}\sigma_0$ is found.}
\end{figure}

\begin{figure}
\includegraphics*{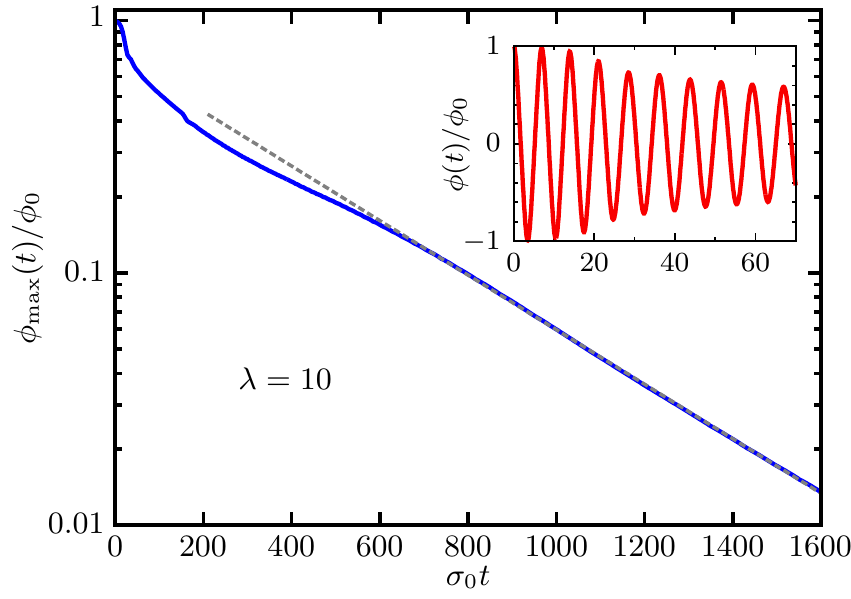}%
\vspace{-1.45ex}
\caption{\label{fig:phi_l10_phi-t}For $\lambda=10$, we observe an exponential decay of the field amplitude $\phi_\mathrm{max}(t)$ with a rate of about $\sim 10^{-3}\sigma_0$ (dashed line).}
\end{figure}

Going beyond the weakly coupled regime, we present the dynamics of the macroscopic field for $\lambda=1\text{ and }10$ in Figs.~\ref{fig:phi_l1_phi-t}%
 and~\ref{fig:phi_l10_phi-t}, %
respectively. Overall, the oscillation frequency and, therefore, the effective mass again remain approximately constant during the evolution. For intermediate times, i.e.\ after the end of the instability and during the approach to the turbulent stage of the evolution, we observe an intermediate power law~$\sim t^{-c}$ with $c\approx0.4$ for the smaller coupling displayed, $\lambda=1$. Later on, the maximum field amplitude follows a very slow exponential decay in contrast to the weak power-law behavior found in the weakly coupled regime. In the case of the stronger coupling of $\lambda=10$, the field oscillations are exponentially damped with a rate of about $\sim 10^{-3}\sigma_0$, which results in an almost vanishing field expectation value for the latest times shown.

\subsection{\label{sec:NonperturbativeInfraredBroken}Nonperturbative Infrared Regime}
\begin{figure}
\vspace{6.95ex}
\includegraphics*{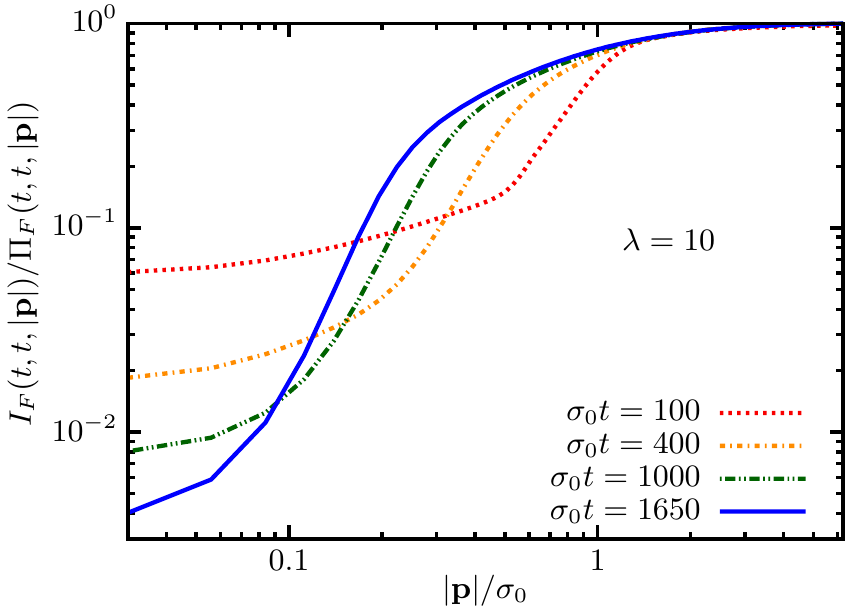}%
\vspace{-0.5ex}
\caption{\label{fig:phi_l10_ratio-p}The ratio~\coloreqref{eq:ratio} for macroscopic field initial conditions with $\lambda=10$ for different times.\vspace{2.6ex}}
\end{figure}

As for the fluctuation initial conditions, we also investigate the ratio~\coloreqref{eq:ratio} in the symmetry-broken regime. We find the same behavior as discussed for the symmetric regime in Sec.~\ref{sec:NonperturbativeInfraredSymmetric} for interaction strengths in the range $\lambda=\numrange{0.01}{1}$. This agreement corresponds to the fact that one observes very similar results from the numerical calculations with both initial conditions in the low-momentum regime in Secs.~\ref{sec:NonthermFixedPointsSymmetric} and~\ref{sec:NonthermalFixedPointsBroken} (cf.~Figs.~\ref{fig:f_l001+01+1_f-p}, \ref{fig:phi_l001_f-p} and~\ref{fig:phi_l1_f-p}). For $\lambda=10$, the ratio is displayed in Fig.~\ref{fig:phi_l10_ratio-p}. %
In view of our findings for fluctuation initial conditions, it exhibits still a remarkably similar behavior to the one observed for weaker couplings.

\section{\label{sec:Summary}Summary and Conclusions}

For the example of a self-interacting $N$-component scalar quantum field theory in a large-$N$ expansion to next-to-leading order, we have shown that nonthermal fixed points are approached from large-field or overoccupied initial conditions. Remarkably, we find the associated transient universal scaling behavior also beyond the weak-coupling limit, well beyond the range of validity of classical-statistical simulations. This concerns in particular the presence of an inverse particle cascade with very large occupancies in the infrared, the characteristic properties of which can be found for couplings as large as $\lambda\sim\mathcal{O}(10)$. While these results may be unexpected in view of genuine weak-coupling techniques such as the classical-statistical approach or quantum kinetic theory, we emphasize that corresponding findings of universal behavior being insensitive to the values of couplings are well known for the (albeit different) case of scaling behavior near phase transitions in thermal equilibrium.

\vskip4pt
The central results of this paper can all be seen in the bottom panel of Fig.~\ref{fig:f_l001+01+1_f-p}, which displays the evolution of the distribution function at $\lambda=1$ starting from overoccupied initial conditions. It demonstrates the turbulent dual cascade picture, with an inverse particle cascade and a direct energy cascade, directly in quantum field theory and beyond the weak-coupling limit. Moreover, it shows the emergence of the expected perturbative scaling exponent $\kappa_\mathrm{E}=5/3$ associated with the direct energy cascade at large momenta for vanishing macroscopic field. It also reflects the fact that at small momenta the dynamics becomes nonrelativistic with a nonperturbative stationary exponent of $\kappa_\mathrm{N}=5$.

In addition, we analyzed the role and importance of~the quantum corrections to stabilize the quantum vacuum beyond the weak-coupling limit and presented a detailed comparison to the dynamics starting from strong-field initial conditions. In scenarios including a macroscopic field for the coupling parameter $\lambda=10$, we discovered a hybrid picture of turbulent behavior reminiscent of the particle cascade in the infrared and an approach towards thermal equilibrium in the ultraviolet for the transverse modes, while the longitudinal degrees of freedom approach quantum thermal equilibrium in the entire momentum range. 

\vskip4pt
Our findings challenge perturbative kinetic descriptions of thermalization dynamics starting from strong-field or overoccupied initial conditions. This has to be seen also in view of the findings of~\colorcite{Berges:2015ixa}, where \mbox{disagreements} between classical-statistical simulations and kinetic theory even in the weak-coupling regime were observed, and where the origin of the discrepancies can be traced back to the emergence of nonperturbatively large occupancies in the infrared. In turn, the successful application of the NLO large-$N$ summation in describing this physics may be used as a starting point for improved kinetic descriptions~\colorcite{Orioli:2015dxa}.
    
The nonrelativistic behavior in the infrared allows also direct comparisons to the dynamics of ultracold quantum gas systems out of equilibrium~\colorcite{Nowak:2011sk}. Based on the universality observed between highly occupied scalar and gauge field dynamics at weak couplings in~\colorcite{Berges:2014bba}, our results may also be helpful in understanding the dynamics of aspects of gauge theories far from equilibrium beyond the weak-coupling limit relevant for relativistic heavy-ion collisions. For further investigations related to cosmological inflation, it should be considered to extend the computations including fermions~\colorcite{Berges:2010zv, Berges:2013oba} to isotropically expanding space-times for more realistic models, which has been started in~\colorcite{Serreau:2013psa, Gautier:2015pca} for de Sitter space.

\begin{acknowledgments}
We would like to thank Kirill Boguslavski, Daniil Gel-fand, Valentin Kasper, Asier Pi{\~n}eiro Orioli, Alexander Rothkopf, S{\"o}ren Schlichting and Raju Venugopalan for discussions/collaboration on related work. B.\,W.\ gratefully acknowledges support from the Studienstiftung des deutschen Volkes, a Starting Grant of the European Research Council (ERC STG Grant No.~279617), a Cambridge European Scholarship of the Cambridge Trust and an STFC Studentship. Parts of this work were performed on the computational resource bwUniCluster funded by the Ministry of Science, Research and the Arts Baden-W{\"u}rttemberg and the Universities of the State of Baden-W{\"u}rttemberg, Germany, within the framework program bwHPC. This work is part of and supported by the DFG Collaborative Research Centre ``SFB 1225 (ISOQUANT)''.
\end{acknowledgments}

\appendix
\section{\label{app:2piEoM}NLO Evolution Equations}	

In this appendix, we review the nonequilibrium evolution equations from the large-$N$ expansion to NLO of the 2PI effective action and refer to the literature for their derivation~\colorcite{Berges:2001fi, Aarts:2002dj}. Starting from the Gaussian initial conditions of Sec.~\ref{sec:initialConditions}, the evolution equations for the quantum-statistical anti-commutator $F$ defined in~\coloreqref{eq:defF} and commutator $\rho$ defined in~\coloreqref{eq:defRho} are
\begin{align}
\left[ \Box_x \delta_{ac} + M^2_{ac}(x) \right] F_{cb}(x,y) = 	&- \int_{t_0}^{x^0}{\!\!\!\!\d z\: \Sigma^\rho_{ac}(x,z) F_{cb}(z,y)} 	\notag	\\
													&+\int_{t_0}^{y^0}{\!\!\!\!\d z\: \Sigma^F_{ac}(x,z) \rho_{cb}(z,y)}\,,			\\
\left[ \Box_x \delta_{ac} + M^2_{ac}(x) \right] \rho_{cb}(x,y)= 	&- \int_{y^0}^{x^0}{\!\!\!\!\d z\: \Sigma^\rho_{ac}(x,z) \rho_{cb}(z,y)}
\end{align}
with shorthand notation $\int_{t_1}^{t_2}{\!\d z} \equiv \int_{t_1}^{t_2}{\!\d z^0 \int{\!\d^3 z}}$ and initial time $t_0$.

For a quantum theory with classical Lagrangian~\coloreqref{eq:Lagrangian} and field rescaling~\coloreqref{eq:fieldRescaling}, the space-time-dependent effective mass term $M^2_{ab}(x)$ is given by
\begin{align}
M^2_{ab}(x;\phi,F)	&= \frac{1}{6N}\left[ F_{cc}(x,x) + \phi^2(x) \right]\, \delta_{ab} 				\notag	\\
				&\phantom{=\ }\,+ \frac{1}{3N}\left[ F_{ab}(x,x) + \phi_a(x)\phi_b(x) \right]\,	.	\label{eq:2PI1/NEffectiveMass}
\end{align}
The real and imaginary parts of the self-energy, $\Sigma^F_{ac}(x,y)$ and $\Sigma^\rho_{ac}(x,y)$, are at NLO:
\begin{align}
\Sigma^F_{ab}(x,y) 	&\!=\! -\frac{1}{3N} \bigg\{\! I_F(x,y) \phi_a(x)\phi_b(y) 						\notag	\\
					&\hskip1.3cm	+ \left[I_F(x,y) + P_F(x,y)\right] F_{ab}(x,y) 						\notag	\\
					&\hskip1.3cm	-\hskip-1pt \left(\hskip-1pt\frac{\lambda}{2}\right)^{\!2}\! \left[I_\rho(x,y) + P_\rho(x,y)\right] \rho_{ab}(x,y) \hskip-1pt\bigg\}\, ,	\\
\Sigma^\rho_{ab}(x,y)	&\!=\! -\frac{1}{3N} \bigg\{\! I_\rho(x,y) \phi_a(x)\phi_b(y) 					\notag	\\
					&\hskip1.3cm	+ \left[I_\rho(x,y) + P_\rho(x,y)\right] F_{ab}(x,y) 				\notag	\\
					&\hskip1.3cm	+ \left[I_F(x,y) + P_F(x,y)\right] \rho_{ab}(x,y) \hskip-1pt\bigg\}\, .
\end{align}
Here, the $\phi$-independent summation functions $I_F$ and $I_\rho$ are defined by
\begin{align}
I_F(x,y)		&= \Pi_F(x,y) -  \int_{t_0}^{x^0}{\!\!\!\!\d z\: I_\rho(x,z) \Pi_F(z,y)} 			\notag				\\
			&\phantom{=\ }	+ \int_{t_0}^{y^0}{\!\!\!\!\d z\: I_F(x,z) \Pi_\rho(z,y)}\, ,	\label{eq:2PI1/NIF}	\\
I_\rho(x,y) 	&= \Pi_\rho(x,y) - \int_{y^0}^{x^0}{\!\!\!\!\d z\: I_\rho(x,z) \Pi_\rho(z,y)}\, ,	\label{eq:2PI1/NIrho}
\end{align}
where we used \mbox{$\Pi_F(x,y;F,\rho) = [F_{ab}(x,y)F_{ab}(x,y)\; - $} $(\lambda/2)^2 \rho_{ab}(x,y)\rho_{ab}(x,y)]/(6N)$ as well as $\Pi_\rho(x,y;F,\rho) = [F_{ab}(x,y) \rho_{ab}(x,y)]/(3N)$ as defined in Secs.~\ref{sec:largeN_NLO} and~\ref{sec:NonperturbativeInfraredSymmetric}. The $\phi$-dependent summation functions $P_F$ and $P_\rho$, which vanish in the symmetric regime ($\phi=0$), are given by
\begin{align}
P_F(x,y) 		&= J_F(x,y) - \int_{t_0}^{x^0}{\!\!\!\! \d z\: I_\rho(x,z) J_F(z,y)} 				\notag				\\
			&\phantom{=\ } + \int_{t_0}^{y^0}{\!\!\!\! \d z\: I_F(x,z) J_\rho(z,y)} \,,		\label{eq:2PI1/NPF}	\\ 
P_\rho(x,y) 	&= J_\rho(x,y) - \int_{y^0}^{x^0}{\!\!\!\!	 \d z\: I_\rho(x,z) J_\rho(z,y)} \,,	\label{eq:2PI1/NPrho}
\end{align}
where we defined
\begin{align}
J_F(x,y)		&= H_F(x,y) - \int_{t_0}^{x^0}{\!\!\!\!\d z\: H_\rho(x,z)I_F(z,y)}	\notag	\\
			&\phantom{=\ } + \int_{t_0}^{y^0}{\!\!\!\!\d z\: H_F(x,z)I_\rho(z,y)}\,,	\\
J_\rho(x,y)	&= H_\rho(x,y) - \int_{y^0}^{x^0}{\!\!\!\!\d z\: H_\rho(x,z) I_\rho(z,y)}
\end{align}
{with \mbox{$H_F(x,y) = \phi_a(x)F_{ab}(x,y)\phi_b(y)/(3N)$} and \mbox{$H_\rho(x,y) = \phi_a(x)\rho_{ab}(x,y)\phi_b(y)/(3N)$}. With these definitions, one observes that $I_{F,\rho}$, $P_{F,\rho}$ and $J_{F,\rho}$ have the same form. Finally, the evolution equation for the macro-\unskip\parfillskip 0pt \par}{\newpage\noindent}scopic field at NLO can also be given in terms of these quantities:
\begin{align}
&\left\{\left[ \Box_x + \frac{1}{6N}\phi^2(x) \right] \delta_{ab} + \left.M_{ab}^2(x)\right|_{\phi=0}\right\} \phi_b(x) 	\notag	\\
&\hskip2.5cm = -\int_{t_0}^{x^0}{\!\!\!\!\d y \left.\Sigma^\rho_{ac}(x,y)\right|_{\phi=0}\phi_b(y)}\, .
\end{align}
To conclude this appendix, we note that the classical-statistical field theory limit corresponds to neglecting the quantum-half in the above evolution equations, i.e.~putting everywhere $\lambda/2$ to zero taking into account the field rescaling~\coloreqref{eq:fieldRescaling}. There are no other changes concerning the form of the evolution equations, however, in the classical theory the field two-point correlator plays the role of the anti-commutator expectation value of the quantum theory and the Poisson bracket replaces the commutator expectation value~\colorcite{Aarts:2001yn, Berges:2007ym}.

\section{\label{app:renormalization}Mass Renormalization}
It turns out that it is sufficient for the employed parameter ranges to consider a mass renormalization which cancels the quadratically divergent mass terms at initial time for a cutoff-regularized theory. We explicitly checked that the renormalization procedure, which we describe in the following, leads to cutoff-insensitive results for the accessible grid sizes. Renormalization of initial-value problems is further discussed in~\colorcite{Garny:2009ni}. We calculate the mass counterterms of the Gaussian theory at initial time iteratively. Because the macroscopic field initial conditions break the $\mathrm{O}(N)$ symmetry, the dressings of the longitudinal and transverse degrees of freedom differ and we consider them separately in that case.

For vanishing renormalized vacuum mass, we determine the initial renormalized in-medium mass $M_{0,\parallel/\perp}(\phi,F)$ by self-consistently solving the mass gap equations involving the macroscopic field and the tadpole contribution,
\begin{align}
M^2_{0,\parallel}(\phi,F)	&= \frac{1}{6N}\Bigg[ 3 \int_\pvec^\Lambda\!\! \Flong(0,0,\pabs) + 3\phi^2(0)							\notag	\\
						&\phantom{= \frac{1}{6N}\Bigg[} + (N-1)\int_\pvec^\Lambda\!\! \Ftrans(0,0,\pabs) \Bigg] + \delta m_\parallel^2\, ,		\\[-2ex]
M^2_{0,\perp}(\phi,F) 		&= \frac{1}{6N}\Bigg[ \int_\pvec^\Lambda\!\! \Flong(0,0,\pabs) + \phi^2(0)								\notag	\\
						&\phantom{= \frac{1}{6N}\Bigg[} + (N+1)\int_\pvec^\Lambda\!\! \Ftrans(0,0,\pabs) \Bigg] + \delta m_\perp^2\, ,
\end{align}
where we used the shorthand notation $\int_\pvec^\Lambda \equiv \int^\Lambda\!\!\frac{\d^3 p}{(2\pi)^3}$. Here, the initial statistical propagators depend on the effective mass as well, cf.~\coloreqref{eq:initialConditionsF}. We choose to cancel the leading quadratic cutoff dependence of the three-dimensional momentum integrals over the initial statistical propagators by the counterterms $\delta m_\parallel^2$ and $\delta m_\perp^2$, which are given by
\begin{align}
\delta m_\parallel^2 	&= -\frac{1}{6N}\Bigg[ \frac{3}{2} \int_\pvec^\Lambda\! \left(\pvec^2+M^2_{0,\parallel}\right)^{\!-1/2} 				\notag	\\
					&\phantom{= -\frac{1}{6N}\Bigg[}\, + \frac{(N-1)}{2} \int_\pvec^\Lambda\! \left(\pvec^2+M^2_{0,\perp}\right)^{\!-1/2}\Bigg]\, , 	\\
\delta m_\perp^2 		&= -\frac{1}{6N}\Bigg[ \frac{1}{2} \int_\pvec^\Lambda\! \left(\pvec^2+M^2_{0,\parallel}\right)^{\!-1/2} 				\notag	\\
					&\phantom{= -\frac{1}{6N}\Bigg[}\,  + \frac{(N+1)}{2}\int_\pvec^\Lambda\! \left(\pvec^2+M^2_{0,\perp}\right)^{\!-1/2}\Bigg]\, .
\end{align}
During the initialization procedure these equations are iteratively solved starting with $M^2_{0,\parallel}=M^2_{0,\perp}=0$ until convergence is achieved. The same counterterms are employed to also cancel the associated divergences in the equations of motion.

\section{\label{app:energyMomentumTensor}Energy-Momentum Tensor}
Since the energy density is a conserved quantity, we can use it to check the numerical accuracy of our calculations and find very good stability. We obtain this quantity as the $T^{00}$-component of the energy-momentum tensor $T^{\mu\nu}$ and consider spatially homogeneous and isotropic systems. It is useful to split the energy density $\epsilon = \lim_{V \rightarrow \infty} (T^{00}/V)$ in its classical and fluctuation parts,
\begin{equation}
\epsilon = \epsilon_\text{class}(t) + \epsilon_\text{fluc}(t)\, .
\end{equation}
The former only depends on the macroscopic field and can be easily calculated from the action. For the massless theory, taking into account the field rescaling~\coloreqref{eq:fieldRescaling}, we get
\begin{equation}
\lambda\epsilon_\text{class}(t) = \frac{1}{2} \dot\phi^2(t) + \frac{1}{4! N} \phi^4(t)\, ,
\end{equation}
where we omitted the spatial derivatives as we assume spatial homogeneity. In the symmetric regime, this quantity vanishes identically.

Employing the $1/N$ expansion of the 2PI effective action at NLO for a cutoff-regularized theory, we obtain for the fluctuation part of the energy density with all quantities given in Fourier space:
\begin{align}
\lambda\epsilon_\text{fluc}(t) =\ &\frac{1}{2} \int_\pvec \! \Big[\!\!\left.\partial_t \partial_{t'} \:\! F_{aa}(t,t';\pabs)\right|_{t=t'} + \pvec^2 F_{aa}(t,t;\pabs)	\notag	\\
							&\hphantom{\frac{1}{2} \int_\pvec \! \Big[}+ \left.M^2_{ab}(t)\right|_{\!F=0} F_{ab}(t,t;\pabs)\Big]						\notag	\\
							&+ \frac{1}{4! N} \left(\int_\pvec{\!F_{aa}(t,t;\pabs)}\right)^{\!2} + \frac{1}{2} \int_\pvec{\!I_F(t,t;\pabs)}					\notag	\\
							&+ \int_\pvec{\left[\frac{1}{2} P_F(t,t;\pabs) - \frac{1}{6N} H_F(t,t;\pabs)\right]}\, ,									\label{eq:quantumEnergyDensity}
\end{align}
where summation over repeated indices is implied. The effective mass term $M^2_{ab}$ is provided in~\coloreqref{eq:2PI1/NEffectiveMass}, whereas the real-space summation functions $I_F$ and $P_F$ can be found in~\coloreqref{eq:2PI1/NIF} and~\coloreqref{eq:2PI1/NPF}, respectively. Under the stated assumptions, we additionally have $H_F(t,t^\prime;\pabs)=\phi_a(t)F_{ab}(t,t^\prime,\pabs)\phi_b(t^\prime)$. Finally, the last integral in~\coloreqref{eq:quantumEnergyDensity} is absent in the symmetric regime and we point out that the energy density does not contain the spectral function as it vanishes at equal times due to the bosonic commutation relations.

\xpatchcmd\bibsection{19}{24.5}{}{}
\xpatchcmd\bibsection{\begingroup}{\vskip5.5pt\begingroup}{}{}
\bibliography{references}

\begin{thebibliography}{41}%
\makeatletter
\providecommand \@ifxundefined [1]{%
 \@ifx{#1\undefined}
}%
\providecommand \@ifnum [1]{%
 \ifnum #1\expandafter \@firstoftwo
 \else \expandafter \@secondoftwo
 \fi
}%
\providecommand \@ifx [1]{%
 \ifx #1\expandafter \@firstoftwo
 \else \expandafter \@secondoftwo
 \fi
}%
\providecommand \natexlab [1]{#1}%
\providecommand \enquote  [1]{``#1''}%
\providecommand \bibnamefont  [1]{#1}%
\providecommand \bibfnamefont [1]{#1}%
\providecommand \citenamefont [1]{#1}%
\providecommand \href@noop [0]{\@secondoftwo}%
\providecommand \href [0]{\begingroup \@sanitize@url \@href}%
\providecommand \@href[1]{\@@startlink{#1}\@@href}%
\providecommand \@@href[1]{\endgroup#1\@@endlink}%
\providecommand \@sanitize@url [0]{\catcode `\\12\catcode `\$12\catcode
  `\&12\catcode `\#12\catcode `\^12\catcode `\_12\catcode `\%12\relax}%
\providecommand \@@startlink[1]{}%
\providecommand \@@endlink[0]{}%
\providecommand \url  [0]{\begingroup\@sanitize@url \@url }%
\providecommand \@url [1]{\endgroup\@href {#1}{\urlprefix }}%
\providecommand \urlprefix  [0]{URL }%
\providecommand \Eprint [0]{\href }%
\providecommand \doibase [0]{http://dx.doi.org/}%
\providecommand \selectlanguage [0]{\@gobble}%
\providecommand \bibinfo  [0]{\@secondoftwo}%
\providecommand \bibfield  [0]{\@secondoftwo}%
\providecommand \translation [1]{[#1]}%
\providecommand \BibitemOpen [0]{}%
\providecommand \bibitemStop [0]{}%
\providecommand \bibitemNoStop [0]{.\EOS\space}%
\providecommand \EOS [0]{\spacefactor3000\relax}%
\providecommand \BibitemShut  [1]{\csname bibitem#1\endcsname}%
\let\auto@bib@innerbib\@empty
%</preamble>
\bibitem [{\citenamefont {{Khlebnikov}}\ and\ \citenamefont
  {{Tkachev}}(1996)}]{Khlebnikov:1996mc}%
  \BibitemOpen
  \bibfield  {author} {\bibinfo {author} {\bibfnamefont {S.}~\bibnamefont
  {{Khlebnikov}}}\ and\ \bibinfo {author} {\bibfnamefont {I.}~\bibnamefont
  {{Tkachev}}},\ }\bibfield  {title} {\enquote {\bibinfo {title} {{Classical
  Decay of the Inflaton}},}\ }\href {\doibase 10.1103/PhysRevLett.77.219}
  {\bibfield  {journal} {\bibinfo  {journal} {Phys. Rev. Lett.}\ }\textbf
  {\bibinfo {volume} {77}},\ \bibinfo {pages} {219} (\bibinfo {year} {1996})},\
  \Eprint {http://arxiv.org/abs/hep-ph/9603378} {arXiv:hep-ph/9603378 [hep-ph]}
  \BibitemShut {NoStop}%
%%CITATION = HEP-PH/9603378;%%
\bibitem [{\citenamefont {{Micha}}\ and\ \citenamefont
  {{Tkachev}}(2003)}]{Micha:2002ey}%
  \BibitemOpen
  \bibfield  {author} {\bibinfo {author} {\bibfnamefont {R.}~\bibnamefont
  {{Micha}}}\ and\ \bibinfo {author} {\bibfnamefont {I.}~\bibnamefont
  {{Tkachev}}},\ }\bibfield  {title} {\enquote {\bibinfo {title} {{Relativistic
  Turbulence: A Long Way from Preheating to Equilibrium}},}\ }\href {\doibase
  10.1103/PhysRevLett.90.121301} {\bibfield  {journal} {\bibinfo  {journal}
  {Phys. Rev. Lett.}\ }\textbf {\bibinfo {volume} {90}},\ \bibinfo {pages}
  {121301} (\bibinfo {year} {2003})},\ \Eprint
  {http://arxiv.org/abs/hep-ph/0210202} {arXiv:hep-ph/0210202 [hep-ph]}
  \BibitemShut {NoStop}%
%%CITATION = HEP-PH/0210202;%%
\bibitem [{\citenamefont {{Berges}}\ \emph {et~al.}(2008)\citenamefont
  {{Berges}}, \citenamefont {{Rothkopf}},\ and\ \citenamefont
  {{Schmidt}}}]{Berges:2008wm}%
  \BibitemOpen
  \bibfield  {author} {\bibinfo {author} {\bibfnamefont {J.}~\bibnamefont
  {{Berges}}}, \bibinfo {author} {\bibfnamefont {A.}~\bibnamefont
  {{Rothkopf}}},\ and\ \bibinfo {author} {\bibfnamefont {J.}~\bibnamefont
  {{Schmidt}}},\ }\bibfield  {title} {\enquote {\bibinfo {title} {{Nonthermal
  Fixed Points: Effective Weak Coupling for Strongly Correlated Systems Far
  from Equilibrium}},}\ }\href {\doibase 10.1103/PhysRevLett.101.041603}
  {\bibfield  {journal} {\bibinfo  {journal} {Phys. Rev. Lett.}\ }\textbf
  {\bibinfo {volume} {101}},\ \bibinfo {pages} {041603} (\bibinfo {year}
  {2008})},\ \Eprint {http://arxiv.org/abs/0803.0131} {arXiv:0803.0131
  [hep-ph]} \BibitemShut {NoStop}%
%%CITATION = ARXIV:0803.0131;%%
\bibitem [{\citenamefont {{Lappi}}\ and\ \citenamefont
  {{McLerran}}(2006)}]{Lappi:2006fp}%
  \BibitemOpen
  \bibfield  {author} {\bibinfo {author} {\bibfnamefont {T.}~\bibnamefont
  {{Lappi}}}\ and\ \bibinfo {author} {\bibfnamefont {L.}~\bibnamefont
  {{McLerran}}},\ }\bibfield  {title} {\enquote {\bibinfo {title} {{Some
  Features of the Glasma}},}\ }\href {\doibase 10.1016/j.nuclphysa.2006.04.001}
  {\bibfield  {journal} {\bibinfo  {journal} {Nucl. Phys. A}\ }\textbf
  {\bibinfo {volume} {772}},\ \bibinfo {pages} {200} (\bibinfo {year}
  {2006})},\ \Eprint {http://arxiv.org/abs/hep-ph/0602189}
  {arXiv:hep-ph/0602189 [hep-ph]} \BibitemShut {NoStop}%
%%CITATION = HEP-PH/0602189;%%
\bibitem [{\citenamefont {{Gelis}}\ \emph {et~al.}(2010)\citenamefont
  {{Gelis}}, \citenamefont {{Iancu}}, \citenamefont {{Jalilian-Marian}},\ and\
  \citenamefont {{Venugopalan}}}]{Gelis:2010nm}%
  \BibitemOpen
  \bibfield  {author} {\bibinfo {author} {\bibfnamefont {F.}~\bibnamefont
  {{Gelis}}}, \bibinfo {author} {\bibfnamefont {E.}~\bibnamefont {{Iancu}}},
  \bibinfo {author} {\bibfnamefont {J.}~\bibnamefont {{Jalilian-Marian}}},\
  and\ \bibinfo {author} {\bibfnamefont {R.}~\bibnamefont {{Venugopalan}}},\
  }\bibfield  {title} {\enquote {\bibinfo {title} {{The Color Glass
  Condensate}},}\ }\href {\doibase 10.1146/annurev.nucl.010909.083629}
  {\bibfield  {journal} {\bibinfo  {journal} {Ann. Rev. Nucl. Part. Sci.}\
  }\textbf {\bibinfo {volume} {60}},\ \bibinfo {pages} {463} (\bibinfo {year}
  {2010})},\ \Eprint {http://arxiv.org/abs/1002.0333} {arXiv:1002.0333
  [hep-ph]} \BibitemShut {NoStop}%
%%CITATION = ARXIV:1002.0333;%%
\bibitem [{\citenamefont {{Berges}}\ \emph
  {et~al.}(2014{\natexlab{a}})\citenamefont {{Berges}}, \citenamefont
  {{Boguslavski}}, \citenamefont {{Schlichting}},\ and\ \citenamefont
  {{Venugopalan}}}]{Berges:2013eia}%
  \BibitemOpen
  \bibfield  {author} {\bibinfo {author} {\bibfnamefont {J.}~\bibnamefont
  {{Berges}}}, \bibinfo {author} {\bibfnamefont {K.}~\bibnamefont
  {{Boguslavski}}}, \bibinfo {author} {\bibfnamefont {S.}~\bibnamefont
  {{Schlichting}}},\ and\ \bibinfo {author} {\bibfnamefont {R.}~\bibnamefont
  {{Venugopalan}}},\ }\bibfield  {title} {\enquote {\bibinfo {title}
  {{Turbulent Thermalization Process in Heavy-Ion Collisions at
  Ultrarelativistic Energies}},}\ }\href {\doibase 10.1103/PhysRevD.89.074011}
  {\bibfield  {journal} {\bibinfo  {journal} {Phys. Rev. D}\ }\textbf {\bibinfo
  {volume} {89}},\ \bibinfo {pages} {074011} (\bibinfo {year}
  {2014}{\natexlab{a}})},\ \Eprint {http://arxiv.org/abs/1303.5650}
  {arXiv:1303.5650 [hep-ph]} \BibitemShut {NoStop}%
%%CITATION = ARXIV:1303.5650;%%
\bibitem [{\citenamefont {{Berges}}\ \emph
  {et~al.}(2014{\natexlab{b}})\citenamefont {{Berges}}, \citenamefont
  {{Boguslavski}}, \citenamefont {{Schlichting}},\ and\ \citenamefont
  {{Venugopalan}}}]{Berges:2013fga}%
  \BibitemOpen
  \bibfield  {author} {\bibinfo {author} {\bibfnamefont {J.}~\bibnamefont
  {{Berges}}}, \bibinfo {author} {\bibfnamefont {K.}~\bibnamefont
  {{Boguslavski}}}, \bibinfo {author} {\bibfnamefont {S.}~\bibnamefont
  {{Schlichting}}},\ and\ \bibinfo {author} {\bibfnamefont {R.}~\bibnamefont
  {{Venugopalan}}},\ }\bibfield  {title} {\enquote {\bibinfo {title}
  {{Universal Attractor in a Highly Occupied Non-Abelian Plasma}},}\ }\href
  {\doibase 10.1103/PhysRevD.89.114007} {\bibfield  {journal} {\bibinfo
  {journal} {Phys. Rev. D}\ }\textbf {\bibinfo {volume} {89}},\ \bibinfo
  {pages} {114007} (\bibinfo {year} {2014}{\natexlab{b}})},\ \Eprint
  {http://arxiv.org/abs/1311.3005} {arXiv:1311.3005 [hep-ph]} \BibitemShut
  {NoStop}%
%%CITATION = ARXIV:1311.3005;%%
\bibitem [{\citenamefont {{Scheppach}}\ \emph {et~al.}(2010)\citenamefont
  {{Scheppach}}, \citenamefont {{Berges}},\ and\ \citenamefont
  {{Gasenzer}}}]{Scheppach:2009wu}%
  \BibitemOpen
  \bibfield  {author} {\bibinfo {author} {\bibfnamefont {C.}~\bibnamefont
  {{Scheppach}}}, \bibinfo {author} {\bibfnamefont {J.}~\bibnamefont
  {{Berges}}},\ and\ \bibinfo {author} {\bibfnamefont {T.}~\bibnamefont
  {{Gasenzer}}},\ }\bibfield  {title} {\enquote {\bibinfo {title} {{Matter-Wave
  Turbulence: Beyond Kinetic Scaling}},}\ }\href {\doibase
  10.1103/PhysRevA.81.033611} {\bibfield  {journal} {\bibinfo  {journal} {Phys.
  Rev. A}\ }\textbf {\bibinfo {volume} {81}},\ \bibinfo {pages} {033611}
  (\bibinfo {year} {2010})},\ \Eprint {http://arxiv.org/abs/0912.4183}
  {arXiv:0912.4183 [cond-mat.quant-gas]} \BibitemShut {NoStop}%
%%CITATION = ARXIV:0912.4183;%%
\bibitem [{\citenamefont {{Nowak}}\ \emph {et~al.}(2012)\citenamefont
  {{Nowak}}, \citenamefont {{Schole}}, \citenamefont {{Sexty}},\ and\
  \citenamefont {{Gasenzer}}}]{Nowak:2011sk}%
  \BibitemOpen
  \bibfield  {author} {\bibinfo {author} {\bibfnamefont {B.}~\bibnamefont
  {{Nowak}}}, \bibinfo {author} {\bibfnamefont {J.}~\bibnamefont {{Schole}}},
  \bibinfo {author} {\bibfnamefont {D.}~\bibnamefont {{Sexty}}},\ and\
  \bibinfo {author} {\bibfnamefont {T.}~\bibnamefont {{Gasenzer}}},\ }\bibfield
   {title} {\enquote {\bibinfo {title} {{Nonthermal Fixed Points, Vortex
  Statistics, and Superfluid Turbulence in an Ultracold Bose Gas}},}\ }\href
  {\doibase 10.1103/PhysRevA.85.043627} {\bibfield  {journal} {\bibinfo
  {journal} {Phys. Rev. A}\ }\textbf {\bibinfo {volume} {85}},\ \bibinfo
  {pages} {043627} (\bibinfo {year} {2012})},\ \Eprint
  {http://arxiv.org/abs/1111.6127} {arXiv:1111.6127 [cond-mat.quant-gas]}
  \BibitemShut {NoStop}%
%%CITATION = ARXIV:1111.6127;%%
\bibitem [{\citenamefont {{Pi{\~n}eiro Orioli}}\ \emph
  {et~al.}(2015)\citenamefont {{Pi{\~n}eiro Orioli}}, \citenamefont
  {{Boguslavski}},\ and\ \citenamefont {{Berges}}}]{Orioli:2015dxa}%
  \BibitemOpen
  \bibfield  {author} {\bibinfo {author} {\bibfnamefont {A.}~\bibnamefont
  {{Pi{\~n}eiro Orioli}}}, \bibinfo {author} {\bibfnamefont {K.}~\bibnamefont
  {{Boguslavski}}},\ and\ \bibinfo {author} {\bibfnamefont {J.}~\bibnamefont
  {{Berges}}},\ }\bibfield  {title} {\enquote {\bibinfo {title} {{Universal
  Self-Similar Dynamics of Relativistic and Nonrelativistic Field Theories Near
  Nonthermal Fixed Points}},}\ }\href {\doibase 10.1103/PhysRevD.92.025041}
  {\bibfield  {journal} {\bibinfo  {journal} {Phys. Rev. D}\ }\textbf {\bibinfo
  {volume} {92}},\ \bibinfo {pages} {025041} (\bibinfo {year} {2015})},\
  \Eprint {http://arxiv.org/abs/1503.02498} {arXiv:1503.02498 [hep-ph]}
  \BibitemShut {NoStop}%
%%CITATION = ARXIV:1503.02498;%%
\bibitem [{\citenamefont {{Berges}}\ \emph
  {et~al.}(2015{\natexlab{a}})\citenamefont {{Berges}}, \citenamefont
  {{Boguslavski}}, \citenamefont {{Schlichting}},\ and\ \citenamefont
  {{Venugopalan}}}]{Berges:2014bba}%
  \BibitemOpen
  \bibfield  {author} {\bibinfo {author} {\bibfnamefont {J.}~\bibnamefont
  {{Berges}}}, \bibinfo {author} {\bibfnamefont {K.}~\bibnamefont
  {{Boguslavski}}}, \bibinfo {author} {\bibfnamefont {S.}~\bibnamefont
  {{Schlichting}}},\ and\ \bibinfo {author} {\bibfnamefont {R.}~\bibnamefont
  {{Venugopalan}}},\ }\bibfield  {title} {\enquote {\bibinfo {title}
  {{Universality Far from Equilibrium: From Superfluid Bose Gases to Heavy-Ion
  Collisions}},}\ }\href {\doibase 10.1103/PhysRevLett.114.061601} {\bibfield
  {journal} {\bibinfo  {journal} {Phys. Rev. Lett.}\ }\textbf {\bibinfo
  {volume} {114}},\ \bibinfo {pages} {061601} (\bibinfo {year}
  {2015}{\natexlab{a}})},\ \Eprint {http://arxiv.org/abs/1408.1670}
  {arXiv:1408.1670 [hep-ph]} \BibitemShut {NoStop}%
%%CITATION = ARXIV:1408.1670;%%
\bibitem [{\citenamefont {{Son}}(1996)}]{Son:1996uv}%
  \BibitemOpen
  \bibfield  {author} {\bibinfo {author} {\bibfnamefont {D.}~\bibnamefont
  {{Son}}},\ }\bibfield  {title} {\enquote {\bibinfo {title} {{Reheating and
  Thermalization in a Simple Scalar Model}},}\ }\href {\doibase
  10.1103/PhysRevD.54.3745} {\bibfield  {journal} {\bibinfo  {journal} {Phys.
  Rev. D}\ }\textbf {\bibinfo {volume} {54}},\ \bibinfo {pages} {3745}
  (\bibinfo {year} {1996})},\ \Eprint {http://arxiv.org/abs/hep-ph/9604340}
  {arXiv:hep-ph/9604340 [hep-ph]} \BibitemShut {NoStop}%
%%CITATION = HEP-PH/9604340;%%
\bibitem [{\citenamefont {{Aarts}}\ and\ \citenamefont
  {{Berges}}(2002)}]{Aarts:2001yn}%
  \BibitemOpen
  \bibfield  {author} {\bibinfo {author} {\bibfnamefont {G.}~\bibnamefont
  {{Aarts}}}\ and\ \bibinfo {author} {\bibfnamefont {J.}~\bibnamefont
  {{Berges}}},\ }\bibfield  {title} {\enquote {\bibinfo {title} {{Classical
  Aspects of Quantum Fields Far from Equilibrium}},}\ }\href {\doibase
  10.1103/PhysRevLett.88.041603} {\bibfield  {journal} {\bibinfo  {journal}
  {Phys. Rev. Lett.}\ }\textbf {\bibinfo {volume} {88}},\ \bibinfo {pages}
  {041603} (\bibinfo {year} {2002})},\ \Eprint
  {http://arxiv.org/abs/hep-ph/0107129} {arXiv:hep-ph/0107129 [hep-ph]}
  \BibitemShut {NoStop}%
%%CITATION = HEP-PH/0107129;%%
\bibitem [{\citenamefont {{Adams}}\ \emph {et~al.}(2013)\citenamefont
  {{Adams}}, \citenamefont {{Chesler}},\ and\ \citenamefont
  {{Liu}}}]{Adams:2012pj}%
  \BibitemOpen
  \bibfield  {author} {\bibinfo {author} {\bibfnamefont {A.}~\bibnamefont
  {{Adams}}}, \bibinfo {author} {\bibfnamefont {P.}~\bibnamefont {{Chesler}}},\
  and\ \bibinfo {author} {\bibfnamefont {H.}~\bibnamefont {{Liu}}},\
  }\bibfield  {title} {\enquote {\bibinfo {title} {{Holographic Vortex Liquids
  and Superfluid Turbulence}},}\ }\href {\doibase 10.1126/science.1233529}
  {\bibfield  {journal} {\bibinfo  {journal} {Science}\ }\textbf {\bibinfo
  {volume} {341}},\ \bibinfo {pages} {368} (\bibinfo {year} {2013})},\ \Eprint
  {http://arxiv.org/abs/1212.0281} {arXiv:1212.0281 [hep-th]} \BibitemShut
  {NoStop}%
%%CITATION = ARXIV:1212.0281;%%
\bibitem [{\citenamefont {{Ewerz}}\ \emph {et~al.}(2015)\citenamefont
  {{Ewerz}}, \citenamefont {{Gasenzer}}, \citenamefont {{Karl}},\ and\
  \citenamefont {{Samberg}}}]{Ewerz:2014tua}%
  \BibitemOpen
  \bibfield  {author} {\bibinfo {author} {\bibfnamefont {C.}~\bibnamefont
  {{Ewerz}}}, \bibinfo {author} {\bibfnamefont {T.}~\bibnamefont {{Gasenzer}}},
  \bibinfo {author} {\bibfnamefont {M.}~\bibnamefont {{Karl}}},\ and\ \bibinfo
  {author} {\bibfnamefont {A.}~\bibnamefont {{Samberg}}},\ }\bibfield  {title}
  {\enquote {\bibinfo {title} {{Nonthermal Fixed Point in a Holographic
  Superfluid}},}\ }\href {\doibase 10.1007/JHEP05(2015)070} {\bibfield
  {journal} {\bibinfo  {journal} {JHEP}\ }\textbf {\bibinfo {volume} {1505}},\
  \bibinfo {pages} {070} (\bibinfo {year} {2015})},\ \Eprint
  {http://arxiv.org/abs/1410.3472} {arXiv:1410.3472 [hep-th]} \BibitemShut
  {NoStop}%
%%CITATION = ARXIV:1410.3472;%%
\bibitem [{\citenamefont {{Epelbaum}}\ and\ \citenamefont
  {{Gelis}}(2011)}]{Epelbaum:2011pc}%
  \BibitemOpen
  \bibfield  {author} {\bibinfo {author} {\bibfnamefont {T.}~\bibnamefont
  {{Epelbaum}}}\ and\ \bibinfo {author} {\bibfnamefont {F.}~\bibnamefont
  {{Gelis}}},\ }\bibfield  {title} {\enquote {\bibinfo {title} {{Role of
  Quantum Fluctuations in a System with Strong Fields: Spectral Properties and
  Thermalization}},}\ }\href {\doibase 10.1016/j.nuclphysa.2011.09.019}
  {\bibfield  {journal} {\bibinfo  {journal} {Nucl. Phys. A}\ }\textbf
  {\bibinfo {volume} {872}},\ \bibinfo {pages} {210} (\bibinfo {year}
  {2011})},\ \Eprint {http://arxiv.org/abs/1107.0668} {arXiv:1107.0668
  [hep-ph]} \BibitemShut {NoStop}%
%%CITATION = ARXIV:1107.0668;%%
\bibitem [{\citenamefont {{Dusling}}\ \emph {et~al.}(2012)\citenamefont
  {{Dusling}}, \citenamefont {{Epelbaum}}, \citenamefont {{Gelis}},\ and\
  \citenamefont {{Venugopalan}}}]{Dusling:2012ig}%
  \BibitemOpen
  \bibfield  {author} {\bibinfo {author} {\bibfnamefont {K.}~\bibnamefont
  {{Dusling}}}, \bibinfo {author} {\bibfnamefont {T.}~\bibnamefont
  {{Epelbaum}}}, \bibinfo {author} {\bibfnamefont {F.}~\bibnamefont {{Gelis}}},\
  and\ \bibinfo {author} {\bibfnamefont {R.}~\bibnamefont {{Venugopalan}}},\
  }\bibfield  {title} {\enquote {\bibinfo {title} {{Instability-Induced
  Pressure Isotropization in a Longitudinally Expanding System}},}\ }\href
  {\doibase 10.1103/PhysRevD.86.085040} {\bibfield  {journal} {\bibinfo
  {journal} {Phys. Rev. D}\ }\textbf {\bibinfo {volume} {86}},\ \bibinfo
  {pages} {085040} (\bibinfo {year} {2012})},\ \Eprint
  {http://arxiv.org/abs/1206.3336} {arXiv:1206.3336 [hep-ph]} \BibitemShut
  {NoStop}%
%%CITATION = ARXIV:1206.3336;%%
\bibitem [{\citenamefont {{Epelbaum}}\ and\ \citenamefont
  {{Gelis}}(2013)}]{Gelis:2013rba}%
  \BibitemOpen
  \bibfield  {author} {\bibinfo {author} {\bibfnamefont {T.}~\bibnamefont
  {{Epelbaum}}}\ and\ \bibinfo {author} {\bibfnamefont {F.}~\bibnamefont
  {{Gelis}}},\ }\bibfield  {title} {\enquote {\bibinfo {title} {{Pressure
  Isotropization in High-Energy Heavy-Ion Collisions}},}\ }\href {\doibase
  10.1103/PhysRevLett.111.232301} {\bibfield  {journal} {\bibinfo  {journal}
  {Phys. Rev. Lett.}\ }\textbf {\bibinfo {volume} {111}},\ \bibinfo {pages}
  {232301} (\bibinfo {year} {2013})},\ \Eprint {http://arxiv.org/abs/1307.2214}
  {arXiv:1307.2214 [hep-ph]} \BibitemShut {NoStop}%
%%CITATION = ARXIV:1307.2214;%%
\bibitem [{\citenamefont {{Berges}}\ \emph
  {et~al.}(2014{\natexlab{c}})\citenamefont {{Berges}}, \citenamefont
  {{Boguslavski}}, \citenamefont {{Schlichting}},\ and\ \citenamefont
  {{Venugopalan}}}]{Berges:2013lsa}%
  \BibitemOpen
  \bibfield  {author} {\bibinfo {author} {\bibfnamefont {J.}~\bibnamefont
  {{Berges}}}, \bibinfo {author} {\bibfnamefont {K.}~\bibnamefont
  {{Boguslavski}}}, \bibinfo {author} {\bibfnamefont {S.}~\bibnamefont
  {{Schlichting}}},\ and\ \bibinfo {author} {\bibfnamefont {R.}~\bibnamefont
  {{Venugo-palan}}},\ }\bibfield  {title} {\enquote {\bibinfo {title} {{Basin of
  Attraction for Turbulent Thermalization and the Range of Validity of
  Classical-Statistical Simulations}},}\ }\href {\doibase
  10.1007/JHEP05(2014)054} {\bibfield  {journal} {\bibinfo  {journal} {JHEP}\
  }\textbf {\bibinfo {volume} {5}},\ \bibinfo {pages} {54} (\bibinfo {year}
  {2014}{\natexlab{c}})},\ \Eprint {http://arxiv.org/abs/1312.5216}
  {arXiv:1312.5216 [hep-ph]} \BibitemShut {NoStop}%
%%CITATION = ARXIV:1312.5216;%%
\bibitem [{\citenamefont {{Berges}}\ \emph
  {et~al.}(2014{\natexlab{d}})\citenamefont {{Berges}}, \citenamefont
  {{Schenke}}, \citenamefont {{Schlichting}},\ and\ \citenamefont
  {{Venugopalan}}}]{Berges:2014yta}%
  \BibitemOpen
  \bibfield  {author} {\bibinfo {author} {\bibfnamefont {J.}~\bibnamefont
  {{Berges}}}, \bibinfo {author} {\bibfnamefont {B.}~\bibnamefont {{Schenke}}},
  \bibinfo {author} {\bibfnamefont {S.}~\bibnamefont {{Schlichting}}},\ and\
  \bibinfo {author} {\bibfnamefont {R.}~\bibnamefont {{Venugopalan}}},\
  }\bibfield  {title} {\enquote {\bibinfo {title} {{Turbulent Thermalization
  Process in High-Energy Heavy-Ion Collisions}},}\ }\href {\doibase
  10.1016/j.nuclphysa.2014.08.103} {\bibfield  {journal} {\bibinfo  {journal}
  {Nucl. Phys. A}\ }\textbf {\bibinfo {volume} {931}},\ \bibinfo {pages} {348}
  (\bibinfo {year} {2014}{\natexlab{d}})},\ \Eprint
  {http://arxiv.org/abs/1409.1638} {arXiv:1409.1638 [hep-ph]} \BibitemShut
  {NoStop}%
%%CITATION = ARXIV:1409.1638;%%
\bibitem [{\citenamefont {{Kurkela}}\ and\ \citenamefont
  {{Zhu}}(2015)}]{Kurkela:2015qoa}%
  \BibitemOpen
  \bibfield  {author} {\bibinfo {author} {\bibfnamefont {A.}~\bibnamefont
  {{Kurkela}}}\ and\ \bibinfo {author} {\bibfnamefont {Y.}~\bibnamefont
  {{Zhu}}},\ }\bibfield  {title} {\enquote {\bibinfo {title} {{Isotropization
  and Hydrodynamization in Weakly Coupled Heavy-Ion Collisions}},}\ }\href
  {\doibase 10.1103/PhysRevLett.115.182301} {\bibfield  {journal} {\bibinfo
  {journal} {Phys. Rev. Lett.}\ }\textbf {\bibinfo {volume} {115}},\ \bibinfo
  {pages} {182301} (\bibinfo {year} {2015})},\ \Eprint
  {http://arxiv.org/abs/1506.06647} {arXiv:1506.06647 [hep-ph]} \BibitemShut
  {NoStop}%
%%CITATION = ARXIV:1506.06647;%%
\bibitem [{\citenamefont {{Epelbaum}}\ \emph {et~al.}(2015)\citenamefont
  {{Epelbaum}}, \citenamefont {{Gelis}}, \citenamefont {{Jeon}}, \citenamefont
  {{Moore}},\ and\ \citenamefont {{Wu}}}]{Epelbaum:2015vxa}%
  \BibitemOpen
  \bibfield  {author} {\bibinfo {author} {\bibfnamefont {T.}~\bibnamefont
  {{Epelbaum}}}, \bibinfo {author} {\bibfnamefont {F.}~\bibnamefont {{Gelis}}},
  \bibinfo {author} {\bibfnamefont {S.}~\bibnamefont {{Jeon}}}, \bibinfo
  {author} {\bibfnamefont {G.}~\bibnamefont {{Moore}}},\ and\ \bibinfo
  {author} {\bibfnamefont {B.}~\bibnamefont {{Wu}}},\ }\bibfield  {title}
  {\enquote {\bibinfo {title} {{Kinetic Theory of a Longitudinally Expanding
  System of Scalar Particles}},}\ }\href {\doibase 10.1007/JHEP09(2015)117}
  {\bibfield  {journal} {\bibinfo  {journal} {JHEP}\ }\textbf {\bibinfo
  {volume} {09}},\ \bibinfo {pages} {117} (\bibinfo {year} {2015})},\ \Eprint
  {http://arxiv.org/abs/1506.05580} {arXiv:1506.05580 [hep-ph]} \BibitemShut
  {NoStop}%
%%CITATION = ARXIV:1506.05580;%%
\bibitem [{\citenamefont {{Berges}}\ \emph
  {et~al.}(2015{\natexlab{b}})\citenamefont {{Berges}}, \citenamefont
  {{Boguslavski}}, \citenamefont {{Schlichting}},\ and\ \citenamefont
  {{Venugopalan}}}]{Berges:2015ixa}%
  \BibitemOpen
  \bibfield  {author} {\bibinfo {author} {\bibfnamefont {J.}~\bibnamefont
  {{Berges}}}, \bibinfo {author} {\bibfnamefont {K.}~\bibnamefont
  {{Boguslavski}}}, \bibinfo {author} {\bibfnamefont {S.}~\bibnamefont
  {{Schlichting}}},\ and\ \bibinfo {author} {\bibfnamefont {R.}~\bibnamefont
  {{Venugopalan}}},\ }\bibfield  {title} {\enquote {\bibinfo {title}
  {{Nonequilibrium Fixed Points in Longitudinally Expanding Scalar Theories:
  Infrared Cascade, Bose Condensation and a Challenge for Kinetic Theory}},}\
  }\href {\doibase 10.1103/PhysRevD.92.096006} {\bibfield  {journal} {\bibinfo
  {journal} {Phys. Rev. D}\ }\textbf {\bibinfo {volume} {92}},\ \bibinfo
  {pages} {096006} (\bibinfo {year} {2015}{\natexlab{b}})},\ \Eprint
  {http://arxiv.org/abs/1508.03073} {arXiv:1508.03073 [hep-ph]} \BibitemShut
  {NoStop}%
%%CITATION = ARXIV:1508.03073;%%
\bibitem [{\citenamefont {{Berges}}(2002)}]{Berges:2001fi}%
  \BibitemOpen
  \bibfield  {author} {\bibinfo {author} {\bibfnamefont {J.}~\bibnamefont
  {{Berges}}},\ }\bibfield  {title} {\enquote {\bibinfo {title} {{Controlled
  Nonperturbative Dynamics of Quantum Fields out of Equilibrium}},}\ }\href
  {\doibase 10.1016/S0375-9474(01)01295-7} {\bibfield  {journal} {\bibinfo
  {journal} {Nucl. Phys. A}\ }\textbf {\bibinfo {volume} {699}},\ \bibinfo
  {pages} {847} (\bibinfo {year} {2002})},\ \Eprint
  {http://arxiv.org/abs/hep-ph/0105311} {arXiv:hep-ph/0105311 [hep-ph]}
  \BibitemShut {NoStop}%
%%CITATION = HEP-PH/0105311;%%
\bibitem [{\citenamefont {{Aarts}}\ \emph {et~al.}(2002)\citenamefont
  {{Aarts}}, \citenamefont {{Ahrensmeier}}, \citenamefont {{Baier}},
  \citenamefont {{Berges}},\ and\ \citenamefont {{Serreau}}}]{Aarts:2002dj}%
  \BibitemOpen
  \bibfield  {author} {\bibinfo {author} {\bibfnamefont {G.}~\bibnamefont
  {{Aarts}}}, \bibinfo {author} {\bibfnamefont {D.}~\bibnamefont
  {{Ahrensmeier}}}, \bibinfo {author} {\bibfnamefont {R.}~\bibnamefont
  {{Baier}}}, \bibinfo {author} {\bibfnamefont {J.}~\bibnamefont {{Berges}}},\
  and\ \bibinfo {author} {\bibfnamefont {J.}~\bibnamefont {{Serreau}}},\
  }\bibfield  {title} {\enquote {\bibinfo {title} {{Far-from-Equilibrium
  Dynamics with Broken Symmetries from the $1/N$ Expansion of the 2PI Effective
  Action}},}\ }\href {\doibase 10.1103/PhysRevD.66.045008} {\bibfield
  {journal} {\bibinfo  {journal} {Phys. Rev. D}\ }\textbf {\bibinfo {volume}
  {66}},\ \bibinfo {pages} {045008} (\bibinfo {year} {2002})},\ \Eprint
  {http://arxiv.org/abs/hep-ph/0201308} {arXiv:hep-ph/0201308 [hep-ph]}
  \BibitemShut {NoStop}%
%%CITATION = HEP-PH/0201308;%%
\bibitem [{\citenamefont {{Berges}}()}]{Berges:2015kfa}%
  \BibitemOpen
  \bibfield  {author} {\bibinfo {author} {\bibfnamefont {J.}~\bibnamefont
  {{Berges}}},\ }\bibfield  {title} {\enquote {\bibinfo {title}
  {{Nonequilibrium Quantum Fields: From Cold Atoms to Cosmology}},}\
  }\href@noop {} {\ }\Eprint {http://arxiv.org/abs/1503.02907}
  {arXiv:1503.02907 [hep-ph]} \BibitemShut {NoStop}%
%%CITATION = ARXIV:1503.02907;%%
\bibitem [{\citenamefont {{Zinn-Justin}}(2002)}]{ZinnJustin:2002ru}%
  \BibitemOpen
  \bibfield  {author} {\bibinfo {author} {\bibfnamefont {J.}~\bibnamefont
  {{Zinn-Justin}}},\ }\href
  {http://ukcatalogue.oup.com/product/9780198509233.do} {\emph {\bibinfo
  {title} {{Quantum Field Theory and Critical Phenomena}}}}\ (\bibinfo
  {publisher} {Oxford Univ.\ Press, Oxford},\ \bibinfo {year}
  {2002})\BibitemShut {NoStop}%
%%CITATION = IMPHA,113,1;%%
\bibitem [{\citenamefont {{Alford}}\ \emph {et~al.}(2004)\citenamefont
  {{Alford}}, \citenamefont {{Berges}},\ and\ \citenamefont
  {{Cheyne}}}]{Alford:2004jj}%
  \BibitemOpen
  \bibfield  {author} {\bibinfo {author} {\bibfnamefont {M.}~\bibnamefont
  {{Alford}}}, \bibinfo {author} {\bibfnamefont {J.}~\bibnamefont {{Berges}}},\
  and\ \bibinfo {author} {\bibfnamefont {J.}~\bibnamefont {{Cheyne}}},\
  }\bibfield  {title} {\enquote {\bibinfo {title} {{Critical Phenomena from the
  Two-Particle Irreducible $1/N$ Expansion}},}\ }\href {\doibase
  10.1103/PhysRevD.70.125002} {\bibfield  {journal} {\bibinfo  {journal} {Phys.
  Rev. D}\ }\textbf {\bibinfo {volume} {70}},\ \bibinfo {pages} {125002}
  (\bibinfo {year} {2004})},\ \Eprint {http://arxiv.org/abs/hep-ph/0404059}
  {arXiv:hep-ph/0404059 [hep-ph]} \BibitemShut {NoStop}%
%%CITATION = HEP-PH/0404059;%%
\bibitem [{\citenamefont {{Berges}}\ and\ \citenamefont
  {{Serreau}}(2003)}]{Berges:2002cz}%
  \BibitemOpen
  \bibfield  {author} {\bibinfo {author} {\bibfnamefont {J.}~\bibnamefont
  {{Berges}}}\ and\ \bibinfo {author} {\bibfnamefont {J.}~\bibnamefont
  {{Serreau}}},\ }\bibfield  {title} {\enquote {\bibinfo {title} {{Parametric
  Resonance in Quantum Field Theory}},}\ }\href {\doibase
  10.1103/PhysRevLett.91.111601} {\bibfield  {journal} {\bibinfo  {journal}
  {Phys. Rev. Lett.}\ }\textbf {\bibinfo {volume} {91}},\ \bibinfo {pages}
  {111601} (\bibinfo {year} {2003})},\ \Eprint
  {http://arxiv.org/abs/hep-ph/0208070} {arXiv:hep-ph/0208070 [hep-ph]}
  \BibitemShut {NoStop}%
%%CITATION = HEP-PH/0208070;%%
\bibitem [{\citenamefont {{Berges}}\ and\ \citenamefont
  {{Hoffmeister}}(2009)}]{Berges:2008sr}%
  \BibitemOpen
  \bibfield  {author} {\bibinfo {author} {\bibfnamefont {J.}~\bibnamefont
  {{Berges}}}\ and\ \bibinfo {author} {\bibfnamefont {G.}~\bibnamefont
  {{Hoffmeister}}},\ }\bibfield  {title} {\enquote {\bibinfo {title}
  {{Nonthermal Fixed Points and the Functional Renormalization Group}},}\
  }\href {\doibase 10.1016/j.nuclphysb.2008.12.017} {\bibfield  {journal}
  {\bibinfo  {journal} {Nucl. Phys. B}\ }\textbf {\bibinfo {volume} {813}},\
  \bibinfo {pages} {383} (\bibinfo {year} {2009})},\ \Eprint
  {http://arxiv.org/abs/0809.5208} {arXiv:0809.5208 [hep-th]} \BibitemShut
  {NoStop}%
%%CITATION = ARXIV:0809.5208;%%
\bibitem [{\citenamefont {{Micha}}\ and\ \citenamefont
  {{Tkachev}}(2004)}]{Micha:2004bv}%
  \BibitemOpen
  \bibfield  {author} {\bibinfo {author} {\bibfnamefont {R.}~\bibnamefont
  {{Micha}}}\ and\ \bibinfo {author} {\bibfnamefont {I.}~\bibnamefont
  {{Tkachev}}},\ }\bibfield  {title} {\enquote {\bibinfo {title} {{Turbulent
  Thermalization}},}\ }\href {\doibase 10.1103/PhysRevD.70.043538} {\bibfield
  {journal} {\bibinfo  {journal} {Phys. Rev. D}\ }\textbf {\bibinfo {volume}
  {70}},\ \bibinfo {pages} {043538} (\bibinfo {year} {2004})},\ \Eprint
  {http://arxiv.org/abs/hep-ph/0403101} {arXiv:hep-ph/0403101 [hep-ph]}
  \BibitemShut {NoStop}%
%%CITATION = HEP-PH/0403101;%%
\bibitem [{\citenamefont {{Berges}}\ and\ \citenamefont
  {{Sexty}}(2011)}]{Berges:2010ez}%
  \BibitemOpen
  \bibfield  {author} {\bibinfo {author} {\bibfnamefont {J.}~\bibnamefont
  {{Berges}}}\ and\ \bibinfo {author} {\bibfnamefont {D.}~\bibnamefont
  {{Sexty}}},\ }\bibfield  {title} {\enquote {\bibinfo {title} {{Strong versus
  Weak Wave-Turbulence in Relativistic Field Theory}},}\ }\href {\doibase
  10.1103/PhysRevD.83.085004} {\bibfield  {journal} {\bibinfo  {journal} {Phys.
  Rev. D}\ }\textbf {\bibinfo {volume} {83}},\ \bibinfo {pages} {085004}
  (\bibinfo {year} {2011})},\ \Eprint {http://arxiv.org/abs/1012.5944}
  {arXiv:1012.5944 [hep-ph]} \BibitemShut {NoStop}%
%%CITATION = ARXIV:1012.5944;%%
\bibitem [{\citenamefont {{Arrizabalaga}}\ \emph {et~al.}(2004)\citenamefont
  {{Arrizabalaga}}, \citenamefont {{Smit}},\ and\ \citenamefont
  {{Tranberg}}}]{Arrizabalaga:2004iw}%
  \BibitemOpen
  \bibfield  {author} {\bibinfo {author} {\bibfnamefont {A.}~\bibnamefont
  {{Arrizabalaga}}}, \bibinfo {author} {\bibfnamefont {J.}~\bibnamefont
  {{Smit}}},\ and\ \bibinfo {author} {\bibfnamefont {A.}~\bibnamefont
  {{Tranberg}}},\ }\bibfield  {title} {\enquote {\bibinfo {title} {{Tachyonic
  Preheating Using 2PI-$1/N$ Dynamics and the Classical Approximation}},}\
  }\href {\doibase 10.1088/1126-6708/2004/10/017} {\bibfield  {journal}
  {\bibinfo  {journal} {JHEP}\ }\textbf {\bibinfo {volume} {10}},\ \bibinfo
  {pages} {017} (\bibinfo {year} {2004})},\ \Eprint
  {http://arxiv.org/abs/hep-ph/0409177} {arXiv:hep-ph/0409177 [hep-ph]}
  \BibitemShut {NoStop}%
%%CITATION = HEP-PH/0409177;%%
\bibitem [{\citenamefont {{Aarts}}\ and\ \citenamefont
  {{Smit}}(1998)}]{Aarts:1997kp}%
  \BibitemOpen
  \bibfield  {author} {\bibinfo {author} {\bibfnamefont {G.}~\bibnamefont
  {{Aarts}}}\ and\ \bibinfo {author} {\bibfnamefont {J.}~\bibnamefont
  {{Smit}}},\ }\bibfield  {title} {\enquote {\bibinfo {title} {{Classical
  Approximation for Time-Dependent Quantum Field Theory: Diagrammatic Analysis
  for Hot Scalar Fields}},}\ }\href {\doibase 10.1016/S0550-3213(97)00723-2}
  {\bibfield  {journal} {\bibinfo  {journal} {Nucl. Phys. B}\ }\textbf
  {\bibinfo {volume} {511}},\ \bibinfo {pages} {451} (\bibinfo {year}
  {1998})},\ \Eprint {http://arxiv.org/abs/hep-ph/9707342}
  {arXiv:hep-ph/9707342 [hep-ph]} \BibitemShut {NoStop}%
%%CITATION = HEP-PH/9707342;%%
\bibitem [{\citenamefont {{Epelbaum}}\ \emph {et~al.}(2014)\citenamefont
  {{Epelbaum}}, \citenamefont {{Gelis}},\ and\ \citenamefont
  {{Wu}}}]{Epelbaum:2014yja}%
  \BibitemOpen
  \bibfield  {author} {\bibinfo {author} {\bibfnamefont {T.}~\bibnamefont
  {{Epelbaum}}}, \bibinfo {author} {\bibfnamefont {F.}~\bibnamefont {{Gelis}}},\
  and\ \bibinfo {author} {\bibfnamefont {B.}~\bibnamefont {{Wu}}},\
  }\bibfield  {title} {\enquote {\bibinfo {title} {{Nonrenormalizability of the
  Classical-Statistical Approximation}},}\ }\href {\doibase
  10.1103/PhysRevD.90.065029} {\bibfield  {journal} {\bibinfo  {journal} {Phys.
  Rev. D}\ }\textbf {\bibinfo {volume} {90}},\ \bibinfo {pages} {065029}
  (\bibinfo {year} {2014})},\ \Eprint {http://arxiv.org/abs/1402.0115}
  {arXiv:1402.0115 [hep-ph]} \BibitemShut {NoStop}%
%%CITATION = ARXIV:1402.0115;%%
\bibitem [{\citenamefont {{Berges}}\ \emph {et~al.}(2011)\citenamefont
  {{Berges}}, \citenamefont {{Gelfand}},\ and\ \citenamefont
  {{Pruschke}}}]{Berges:2010zv}%
  \BibitemOpen
  \bibfield  {author} {\bibinfo {author} {\bibfnamefont {J.}~\bibnamefont
  {{Berges}}}, \bibinfo {author} {\bibfnamefont {D.}~\bibnamefont {{Gelfand}}},\
  and\ \bibinfo {author} {\bibfnamefont {J.}~\bibnamefont {{Pruschke}}},\
  }\bibfield  {title} {\enquote {\bibinfo {title} {{Quantum Theory of Fermion
  Production after Inflation}},}\ }\href {\doibase
  10.1103/PhysRevLett.107.061301} {\bibfield  {journal} {\bibinfo  {journal}
  {Phys. Rev. Lett.}\ }\textbf {\bibinfo {volume} {107}},\ \bibinfo {pages}
  {061301} (\bibinfo {year} {2011})},\ \Eprint {http://arxiv.org/abs/1012.4632}
  {arXiv:1012.4632 [hep-ph]} \BibitemShut {NoStop}%
%%CITATION = ARXIV:1012.4632;%%
\bibitem [{\citenamefont {{Berges}}\ \emph
  {et~al.}(2014{\natexlab{e}})\citenamefont {{Berges}}, \citenamefont
  {{Gelfand}},\ and\ \citenamefont {{Sexty}}}]{Berges:2013oba}%
  \BibitemOpen
  \bibfield  {author} {\bibinfo {author} {\bibfnamefont {J.}~\bibnamefont
  {{Berges}}}, \bibinfo {author} {\bibfnamefont {D.}~\bibnamefont {{Gelfand}}},\
  and\ \bibinfo {author} {\bibfnamefont {D.}~\bibnamefont {{Sexty}}},\
  }\bibfield  {title} {\enquote {\bibinfo {title} {{Amplified Fermion
  Production from Overpopulated Bose Fields}},}\ }\href {\doibase
  10.1103/PhysRevD.89.025001} {\bibfield  {journal} {\bibinfo  {journal} {Phys.
  Rev. D}\ }\textbf {\bibinfo {volume} {89}},\ \bibinfo {pages} {025001}
  (\bibinfo {year} {2014}{\natexlab{e}})},\ \Eprint
  {http://arxiv.org/abs/1308.2180} {arXiv:1308.2180 [hep-ph]} \BibitemShut
  {NoStop}%
%%CITATION = ARXIV:1308.2180;%%
\bibitem [{\citenamefont {{Serreau}}\ and\ \citenamefont
  {{Parentani}}(2013)}]{Serreau:2013psa}%
  \BibitemOpen
  \bibfield  {author} {\bibinfo {author} {\bibfnamefont {J.}~\bibnamefont
  {{Serreau}}}\ and\ \bibinfo {author} {\bibfnamefont {R.}~\bibnamefont
  {{Parentani}}},\ }\bibfield  {title} {\enquote {\bibinfo {title}
  {{Nonperturbative Resummation of De Sitter Infrared Logarithms in the
  Large-$N$ Limit}},}\ }\href {\doibase 10.1103/PhysRevD.87.085012} {\bibfield
  {journal} {\bibinfo  {journal} {Phys. Rev. D}\ }\textbf {\bibinfo {volume}
  {87}},\ \bibinfo {pages} {085012} (\bibinfo {year} {2013})},\ \Eprint
  {http://arxiv.org/abs/1302.3262} {arXiv:1302.3262 [hep-th]} \BibitemShut
  {NoStop}%
%%CITATION = ARXIV:1302.3262;%%
\bibitem [{\citenamefont {{Gautier}}\ and\ \citenamefont
  {{Serreau}}(2015)}]{Gautier:2015pca}%
  \BibitemOpen
  \bibfield  {author} {\bibinfo {author} {\bibfnamefont {F.}~\bibnamefont
  {{Gautier}}}\ and\ \bibinfo {author} {\bibfnamefont {J.}~\bibnamefont
  {{Serreau}}},\ }\bibfield  {title} {\enquote {\bibinfo {title} {{Scalar Field
  Correlator in De Sitter Space at Next-to-Leading Order in a $1/N$
  Expansion}},}\ }\href {\doibase 10.1103/PhysRevD.92.105035} {\bibfield
  {journal} {\bibinfo  {journal} {Phys. Rev. D}\ }\textbf {\bibinfo {volume}
  {92}},\ \bibinfo {pages} {105035} (\bibinfo {year} {2015})},\ \Eprint
  {http://arxiv.org/abs/1509.05546} {arXiv:1509.05546 [hep-th]} \BibitemShut
  {NoStop}%
%%CITATION = ARXIV:1509.05546;%%
\bibitem [{\citenamefont {{Berges}}\ and\ \citenamefont
  {{Gasenzer}}(2007)}]{Berges:2007ym}%
  \BibitemOpen
  \bibfield  {author} {\bibinfo {author} {\bibfnamefont {J.}~\bibnamefont
  {{Berges}}}\ and\ \bibinfo {author} {\bibfnamefont {T.}~\bibnamefont
  {{Gasenzer}}},\ }\bibfield  {title} {\enquote {\bibinfo {title} {{Quantum
  Versus Classical Statistical Dynamics of an Ultracold Bose Gas}},}\ }\href
  {\doibase 10.1103/PhysRevA.76.033604} {\bibfield  {journal} {\bibinfo
  {journal} {Phys. Rev. A}\ }\textbf {\bibinfo {volume} {76}},\ \bibinfo
  {pages} {033604} (\bibinfo {year} {2007})},\ \Eprint
  {http://arxiv.org/abs/cond-mat/0703163} {arXiv:cond-mat/0703163
  [cond-mat.other]} \BibitemShut {NoStop}%
%%CITATION = COND-MAT/0703163;%%
\bibitem [{\citenamefont {{Garny}}\ and\ \citenamefont
  {{M\"uller}}(2009)}]{Garny:2009ni}%
  \BibitemOpen
  \bibfield  {author} {\bibinfo {author} {\bibfnamefont {M.}~\bibnamefont
  {{Garny}}}\ and\ \bibinfo {author} {\bibfnamefont {M.}~\bibnamefont
  {{M\"uller}}},\ }\bibfield  {title} {\enquote {\bibinfo {title}
  {{Kadanoff-Baym Equations with Non-Gaussian Initial Conditions: The
  Equilibrium Limit}},}\ }\href {\doibase 10.1103/PhysRevD.80.085011}
  {\bibfield  {journal} {\bibinfo  {journal} {Phys. Rev. D}\ }\textbf {\bibinfo
  {volume} {80}},\ \bibinfo {pages} {085011} (\bibinfo {year} {2009})},\
  \Eprint {http://arxiv.org/abs/0904.3600} {arXiv:0904.3600 [hep-ph]}
  \BibitemShut {NoStop}%
%%CITATION = ARXIV:0904.3600;%%
\end{thebibliography}%
\end{document}